\documentclass[showpacs,amssymb,onecolumn,floatfix,aps,notitlepage]{revtex4-1}

\usepackage{amssymb,amsfonts,amsmath,bm}
\usepackage{graphics,graphicx}
\usepackage{xr} 

\begin{document}

\title{Strong-coupling theory of counterions between symmetrically 
charged walls: from crystal to fluid phases}

\author{Ladislav \v{S}amaj$^1$} 
\author{Martin Trulsson$^2$}
\author{Emmanuel Trizac$^3$} 

\affiliation{
$^1$Institute of Physics, Slovak Academy of Sciences, Bratislava, Slovakia \\
$^2$Theoretical Chemistry, Lund University, Lund, Sweden\\
$^3$LPTMS, CNRS, Univ. Paris-Sud, Universit\'e Paris-Saclay, 91405 Orsay, France }

\date{\today} 

\begin{abstract}
We study thermal equilibrium of classical pointlike counterions confined 
between symmetrically charged walls at distance $d$.
At very large couplings when the counterion system is in its crystal
phase, a harmonic expansion of particle deviations is made around the
bilayer positions, with a free lattice parameter determined 
from a variational approach.
For each of the two walls, the harmonic expansion implies an effective 
one-body potential at the root of all observables of interest in our 
Wigner Strong-Coupling expansion.
Analytical results for the particle density profile and the pressure 
are in good agreement with numerical Monte Carlo data, for small as well
as intermediate values of $d$ comparable with the Wigner lattice spacing.
While the strong-coupling theory is extended to the fluid regime by using 
the concept of a correlation hole, the Wigner calculations appear 
trustworthy for all electrostatic couplings investigated. 
Our results significantly extend the range of accuracy of analytical 
equations of state for strongly interacting charged planar interfaces.
\end{abstract}

\maketitle

\renewcommand{\theequation}{1.\arabic{equation}}
\setcounter{equation}{0}

\section{Introduction} \label{Sec.intro}
Large macromolecules such as colloids, immersed in polar solvents, are endowed
with a surface density due to the release of bound ions, or the uptake of 
charged species. 
This exchange with the solution, together with the auto-protolysis of water 
in the case of aqueous solvents leads to a solution containing micro-ions of 
both signs. 
However, it is possible to approach the deionized limit where in addition 
to the colloids, the only charged species are counterions of opposite charge.  
The corresponding idealized ``counterions only'' (salt-free)
case does describe well some experiments (see e.g.  \cite{Palberg04}),
and furthermore, it is a useful and often advocated workbench for 
theoretical purposes, be they analytical or computational.
In thermal equilibrium, the equation of state of salt-free models that 
we concentrate on in this work depends on the only free parameter, 
namely the coupling constant $\Xi$ to be defined below.
Such simplified models help us to understand the limiting weak-coupling (WC) 
and strong-coupling (SC) regimes of general Coulomb systems, and can be useful 
as a starting point in specific approaches to charged systems with salt. 

The curved surface of large macromolecules can be replaced by an
infinite plane in the first approximation.
The counterions can be considered as identical classical (i.e., non-quantum)
pointlike particles interacting via the three-dimensional Coulomb potential.
The charged surface and surrounding counterions form in thermal equilibrium 
a neutral electric double layer, see reviews 
\cite{Attard96,Hansen00,Levin02,Messina09}.
The geometry of two parallel equivalently-charged walls with counterions 
in between provides the prototypical study of the effective interaction 
between like-charged macromolecules.
At large enough electrostatic coupling, like-charged colloids can attract
each other, as was shown in experiments 
\cite{Khan85,Kjellander88,Bloomfield91,Rau92,Kekicheff93,Dubois98} 
as well as in numerical simulations 
\cite{Gulbrand84,Kjellander84,Bratko86,Gronbech97,Linse99,LinseLobaskin2000}.
Like-charge attraction explains phenomena like the formation of DNA
condensates \cite{Bloomfield96} and colloidal aggregates \cite{Linse99}.
On the other hand, like-charge attraction is precluded at small couplings, 
unless the microions acquire an internal structure \cite{May08,Kim08}.

The WC limit of Coulomb fluids is described by the Poisson-Boltzmann (PB) 
mean-field theory \cite{LobaskinLinseJCP1999,Andelman06}. 
For systems with counterions only, the PB theory can be viewed as the leading 
term in a systematic loop-expansion \cite{Attard88}. 
The characteristic inverse-power-law form of mean-field results should hold 
exactly for the particle density profile at asymptotically large distances 
from one wall or the pressure for parallel walls at large distances 
\cite{Shklovskii99,Chen06,Santos09}.

In the opposite SC limit, one needs to make a distinction between 
the crystal and fluid regimes.
For infinite and extremely large couplings $\Xi$, the counterions organize 
themselves into a crystal phase \cite{BausHansen80}.
In the absence of dielectric wall images, according to Earnshaw's theorem 
\cite{Earnshaw1842} the counterions stick on the wall surfaces in 
the ground state (infinite coupling).
For one-wall geometry, they form a two-dimensional (2D) hexagonal, 
or equilateral triangular, Wigner crystal. 
In the case of two parallel walls, five distinct (staggered) Wigner bilayers 
were detected as the distance between the walls increases from zero to infinity 
\cite{Falko94,Esfarjani95,Goldoni96,Schweigert99,Weis01,Messina03,Lobaskin07}. 
The controversial aspects of the topic and the critical properties of
the second-order phase transitions were revisited in Ref. \cite{Samaj12a} 
by using an analytic approach based on an expansion of the energy of 
the five structures in generalized Misra functions \cite{Misra}.
The same problem, with asymmetrically charged walls, displays baffling 
complexity \cite{Moritz16}.
In the presence of repulsive dielectric images, the ground-state Wigner 
layers (in the one-wall problem) and bilayers (in the two-walls problem) 
are localized at specific distances from the walls \cite{Samaj12b}.
The ground-state bilayer Wigner crystal played a key role in the first 
theoretical attempts to construct a SC theory valid for large couplings
\cite{Rouzina96,Shklovskii99,Perel99}. 

The classical bilayer Wigner crystal is stable only at extremely 
large couplings.
The problem of its melting to a fluid was studied within the harmonic 
approximation in Refs. \cite{Goldoni96,Schweigert99} where the charged 
particles were allowed to deviate around their ground-state positions only 
along the 2D Wigner planes. 
To describe the SC limit of the fluid phase, a field-theoretic treatment 
was proposed in \cite{Moreira00,Netz01,Moreira02}, in the form of a virial 
(fugacity) expansion. 
While this approach yields the correct leading SC order in the form of 
a single-particle result, it does not capture the right correction to 
leading behavior \cite{Santangelo06,Samaj11}.
The single-particle SC theory was extended to general dielectric walls
\cite{Kanduc07}, asymmetrically charged walls \cite{Kanduc08,Paillusson11},
and Coulomb models with salt \cite{Kanduc10}.
For the one-wall geometry, an attempt was made in Ref. \cite{Burak04}
to construct a universal theory which is operational for an arbitrary coupling.
The theory is based on a mean-field approach to the response of counterions 
to the presence of a test charge.
Other attempts were put forward in 
\cite{Chen06,Rouzina96,Nordholm84,Santangelo06,Hatlo10},
discriminating short and long distance components of Coulomb forces. 
These interesting approaches do not yield analytical and explicit results,
which are our core interest in this paper.

On the analytical side, a strong-coupling theory dwelling on the Wigner 
crystallization was proposed in Ref. \cite{Samaj11}, hereafter referred to
as the Wigner strong-coupling (WSC) approach.
It is based on the harmonic approximation for particle deviations from their
ground-state positions in the Wigner layer or bilayer crystal, along all 
directions.
The leading order turns out to be identical to the virial single-particle 
theory. 
The first correction to the particle density profile is much stronger than 
within the virial view, and in excellent agreement with Monte Carlo (MC) data 
\cite{Samaj11}.
Although the method starts from the existence of a Wigner crystal, 
it works surprisingly well also for intermediate and relatively small 
couplings when the counterion system is in its fluid phase \cite{Samaj11}. 
The rationale behind such an agreement is that the precise structure of ions 
at the plate is not essential, except from the fact that it is strongly 
modulated. 
Hence the success of simplifying theories relying on a correlation hole, 
that can lead to accurate density profiles, up to relatively small coupling 
constants \cite{Samaj16}. 
The idea can even be formulated in conjunction with a test-particle
approach, to yield a self-consistent theory that has the property to be exact 
at both vanishing and infinite couplings \cite{Ivan}.

In this paper, we restrict ourselves to the geometry of two parallel
symmetrically charged walls with no image charges, at distance $d$. 
Our main goal is to derive the equation of state of the system 
(inter-plate pressure), significantly extending the $d$-range where 
analytical results are known.
Indeed, the virial route of \cite{Moreira00,Netz01,Moreira02} yields 
the dominant small-$d$ pressure, and holds at small distances (less than 
the so-called Gouy-Chapman length). 
The subleading correction was computed in  \cite{Samaj11}, with still
a resulting domain of validity limited to very small $d$. 
Here, we show that the definition of an effective one-body 
potential for each of the two walls allows to extend the affordable $d$ range
up to the typical counterion-counterion separation. 
This represents a gain of a factor $\sqrt{\Xi}$ in the distance-range, 
an appreciable improvement. 
To this end, structural vibrations are taken in full in the 
present WSC approach, without any restriction on the distance 
between the two walls. 
Here, it should be kept in mind that at even larger distances, the mean-field 
PB theory takes over and inter-plate pressures are described accordingly 
\cite{Netz01,Chen06,Shklovskii99}.

Our technique is first put to work for very large values of the coupling 
constant, when the system stays in its crystal phase.
The original approaches considering only vibrations along the Wigner surfaces
\cite{Goldoni96,Schweigert99} were based on the harmonic expansions around
the {\em ground-state} Wigner structure.
Here, we leave the characteristic lattice parameter of the Wigner structure 
(around which the harmonic expansion is made) as free; it is determined 
variationally at the end of the calculations, minimizing the free energy. 
Thus the form of the Wigner bilayer depends not only on $d$, but also on 
the coupling constant $\Xi$; such a scenario is confirmed 
qualitatively as well as quantitatively by numerical simulations.
As concerns the fluid phase at large and intermediate values of the coupling
constant, and following similar lines as Ref. \cite{Samaj16}, we relinquish 
the crystal to invoke a correlation hole when calculating the effective 
one-body potential acting on particles close to each of the two walls.
As before, the analytic results for the particle density profile and 
the pressure agree with numerical data up to intermediate inter-wall distances.

The paper is organized as follows.
The definition of the model and a review of its ground-state features 
are presented in Sec. \ref{Sec.model}.
The numerical Monte Carlo method is discussed in Sec. \ref{Sec.Monte}.
Sec. \ref{Sec.crystal} concerns the large-coupling description of 
the crystal phase.
We start by the harmonic expansion of deviations from the crystal positions
in \ref{Subsec.1}, continue by the leading WSC order and the first correction 
of the corresponding thermodynamics (\ref{Subsec.2}) and then consider 
the particle density profile (\ref{Subsec.3}).
The pressure is obtained in two ways: from the thermodynamic route and 
by using the contact theorem. 
Comparison with the numerical results is given in Sec. \ref{Subsec.4}.
The correlation-hole SC approach to the fluid phase is constructed
in Sec. \ref{Sec.fluid}. 
We conclude in Sec. \ref{Sec.conclusion} with a short summary and 
future plans. 

\renewcommand{\theequation}{2.\arabic{equation}}
\setcounter{equation}{0}

\section{Model and its ground state} \label{Sec.model}
In 3D space of points ${\bf r}=(x,y,z)$, we consider two parallel 
walls (plates) at distance $d$, say plate $\Sigma_1$ at $z=0$ and plate 
$\Sigma_2$ at $z=d$.
The plate surfaces $\vert \Sigma_1 \vert = \vert \Sigma_2 \vert = S$ 
along the $(x,y)$ plane are taken as infinite.
The space between the plates will be denoted by
$\Lambda = \{ {\bf r}; 0\le z\le d \}$.
The plate surfaces carry the same fixed homogeneous surface charge density 
$\sigma e$, where $e$ is the elementary charge and say $\sigma>0$.
The electric field due to the charged plates is equal to 0 in the space
between the plates. 
There are $N$ mobile particles constrained to $\Lambda$, for simplicity 
with unit charge $-e$, coined as ``counterions''.
The system as a whole is electro-neutral, i.e. $N = 2\sigma S$.
The particles are immersed in a solution of dielectric constant $\epsilon$,
the dielectric constant of the walls is considered to be the same 
$\epsilon_{\rm w}=\epsilon$, so there are no image forces acting on particles.
In Gaussian units, the charged plates and particles interact pairwisely by 
the 3D Coulomb potential $1/(\epsilon r)$. 

At zero temperature, the particles organize themselves into 
a Wigner crystal structure with the minimal interaction energy.
According to the Earnshaw theorem \cite{Earnshaw1842}, a classical system of
point charges in a domain, which is under the action of direct (not image)
electrostatic forces, cannot be in an equilibrium position, i.e. the charges
stick to the domain's boundary.
In our symmetric case, taking $N$ as an even number, $N/2$ particles 
$i=1,\ldots,N/2$ stick on plate $\Sigma_1$ and the remaining $N/2$ 
particles $i=N/2+1,\ldots,N$ stick on plate $\Sigma_2$.

\begin{figure}[tbp]
\includegraphics[width=0.25\textwidth,clip]{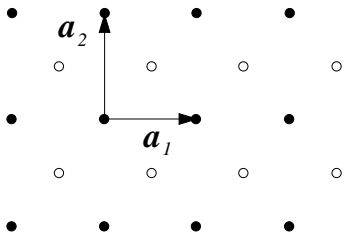}
\caption{Ground-state structures I, II and III (corresponding to different
lattice vectors $\bm{a}_1$ and $\bm{a}_2$) of counterions on two 
equivalently charged plates. 
Open and filled symbols correspond to rectangular positions of particles
on the opposite surfaces. 
The aspect ratio is defined as $\Delta=|{\bf a}_2|/ |{\bf a}_1|$.}
\label{fig:Structures}
\end{figure}

Depending on the dimensionless distance between the plates
\begin{equation}
\eta = d \sqrt{\sigma},
\end{equation}
five distinct bilayer Wigner structures were detected by numerical simulations
\cite{Falko94,Esfarjani95,Goldoni96,Schweigert99,Weis01,Messina03,Lobaskin07}.
In this paper, we study such intervals of $\eta$-values where the staggered
rectangular structures I--III prevail, see Fig. \ref{fig:Structures}. 
A single layer of these structures corresponds to a rectangular lattice 
with the aspect ratio $\Delta$, defined by the primitive translation vectors 
\begin{equation} \label{struct}
\bm{a}_1 = a(1,0) , \qquad \bm{a}_2 = a(0,\Delta) , \qquad
a = \frac{1}{\sqrt{\sigma\Delta}} .
\end{equation} 
The lattice spacing $a$ is determined by the electroneutrality condition
that the surface charge of a rectangle $(e \sigma)a^2\Delta$ must compensate
the charge $-e$ of just one particle per rectangle.
The identical structures on the two plates are shifted with respect to 
one another by a half period $(\bm{a}_1 + \bm{a}_2)/2$.
The position vectors of the particles $i=1,\ldots,N/2$ on the Wigner 
rectangular structure at plate $\Sigma_1$ will be denoted by 
\begin{equation} \label{partplate1}
{\bf r}_i^0 = (a i_x,\Delta a i_y,0) ,
\end{equation}
where $i_x,i_y$ run over all integers; the particle assignment 
$i\to (i_x,i_y)$ is obvious.
Similarly, the position vectors of the particles $i=N/2,\ldots,N$ on the Wigner 
rectangular structure at plate $\Sigma_2$ are denoted by 
\begin{equation} \label{partplate2}
{\bf r}_i^0 = \left( a \left( i_x-\frac{1}{2}\right),
\Delta a \left( i_y-\frac{1}{2}\right),d \right) ,
\end{equation}
where $i_x,i_y$ run again over all integers.
Structure I with $\Delta=\sqrt{3}$ arises naturally in the single-layer
limit $\eta\to 0$ which is known to be characterized by a hexagonal
(equilateral triangular) lattice.
The aspect ratio is from the interval $1<\Delta<\sqrt{3}$ for soft structure 
II and $\Delta=1$ for structure III, i.e. the staggered square lattice.
The phase transformation I--II, which is not a phase transition, takes place
just at $\eta=0$ \cite{Messina03,Samaj12a} or, in other words, 
structure I exists only at $\eta=0$.
The phase transition between structures II and III at $\eta\sim 0.263$ is 
of second order, with singularities of mean-field type \cite{Samaj12a}. 
Phase III has the lowest energy up to $\eta\sim 0.621$.

For all three structures I-III, the energy per particle, 
$e_0=E_0/N$, is expressed as the lattice summation (\ref{e0lattice})
in section \ref{app:B} of the Supplementary Information. 
Writing
\begin{equation} \label{e0}
e_0(\eta,\Delta) = \frac{e^2 \sqrt{\sigma}}{\epsilon} 
\frac{1}{2\sqrt{\pi}} \Sigma(\eta,\Delta)  
\end{equation}
and using techniques introduced in Ref. \cite{Samaj12a}, the function
$\Sigma(\eta,\Delta)$ can be written as an integral over certain products
of Jacobi theta functions, see Eq. (\ref{Sigma}), and subsequently as 
an infinite series of the generalized Misra functions
\begin{equation} \label{znu}
z_{\nu}(x,y)=\int_0^{1/\pi} \frac{{\rm d}t}{t^{\nu}}{\rm e}^{-xt}{\rm e}^{-y/t} ,
\end{equation}
see Eq. (\ref{sigmaMisra}). 
Note that the ordinary Misra functions correspond to $x=0$ \cite{Misra}.
The first few generalized Misra functions $z_{\nu}(x,y)$ with half-integer 
$\nu$-indices are expressed in terms of the complementary error function in 
Eq. (\ref{zerror}) of section \ref{app:A} in the Supplementary Information.
This permits us to use very effectively symbolic softwares.
The series in generalized Misra functions are rapidly converging;
for the well known $\eta=0$ case of the hexagonal lattice with 
$\Delta=\sqrt{3}$, the truncation of the series over $j,k$ at $M=1,2,3,4$ 
reproduces the exact value of the Madelung constant up to $2,5,10,17$ 
decimal digits, respectively \cite{Samaj12a}.
In the present calculations, to keep a high accuracy of the results we truncate 
the series of the generalized Misra functions at $M=6$.
The evaluation of a series takes a fraction of second of CPU on a standard PC.

For a given dimensionless inter-plate distance $\eta$, the actual value of 
the aspect ratio $\Delta$ is determined by the energy minimization condition
\begin{equation} \label{Delta0}
\frac{\partial}{\partial\Delta} e_0(\eta,\Delta) = 0 . 
\end{equation}
This condition determines the dependence $\Delta_0(\eta)$ where the lower
index $0$ means ``in the ground-state'' or, equivalently, at infinite coupling.

\renewcommand{\theequation}{3.\arabic{equation}}
\setcounter{equation}{0}

\section{Monte Carlo simulations} \label{Sec.Monte}
Let the system be in thermal equilibrium at some inverse temperature 
$\beta = 1/(k_{\rm B}T)$.
There are two relevant length scales at nonzero temperature.
The Bjerrum length $\ell_{\rm B} = \beta e^2 /\epsilon$ is the distance at 
which two unit charges interact with thermal energy $k_{\rm B}T$. 
The potential energy of a unit charge at distance $z$ from an isolated wall 
with the surface charge density $e\sigma$ is equal to 
$2\pi e^2\sigma z/\epsilon$.
The unit charge has the potential energy equal to thermal energy $k_{\rm B}T$
at distance from the wall
\begin{equation}
\mu = \frac{1}{2\pi\ell_{\rm B}\sigma} ,
\end{equation}
known as the Gouy-Chapman length.
Since this length is the relevant scale in the direction perpendicular to 
the surfaces of the two walls, the coordinate $z$ will be usually expressed 
in units of $\mu$, $\widetilde{z}=z/\mu$.
The dimensionless coupling parameter $\Xi$, quantifying the strength
of electrostatic correlations, is defined as the ratio
\begin{equation}
\Xi = \frac{\ell_{\rm B}}{\mu} = 2\pi \ell_{\rm B}^2 \sigma .
\end{equation}
The SC regime $\Xi\gg 1$ corresponds to either ``low temperatures'', large 
surface charge densities, or equivalently small dielectric constant.
The lattice spacing of the Wigner structure $a$, which is the characteristic 
length scale in the longitudinal $(x,y)$ plane, is much 
larger than $\mu$ in the SC regime, $a/\mu\propto \sqrt{\Xi}$. 

MC simulations were carried out in a quasi-2D slab geometry for coupling 
parameters ranging between $\Xi=17.5$ and $175000$, where 512 point 
charges were confined between two uniformly charged and flat surfaces, each 
with a surface charged density of $\sigma e$, at various fixed separations $d$. 
The system was periodic in all directions with an extra vacuum slab in 
the $z$-direction perpendicular to the surfaces and between the slab images.
This set-up allowed us to use standard 3D Ewald summation techniques 
to handle the long-ranged electrostatic interactions, with only 
minor re-adaptions to correct for the quasi-2D-dimensionality 
and extra vacuum space \cite{Berkowitz,Mazars}. 
We verified that our vacuum slab is sufficiently wide (typically much wider 
than the separation $d$ between the walls) so as not to influence the results. 
New MC configurations were either generated by trial displacement 
of the point charges or by volume preserving floppy-box moves. 
Two floppy-box moves were utilized: shear or combined biaxial 
compression/decompression (compression along one axis and decompression 
along the other such to preserve the box volume). 
Both deformations were performed in the $(x,y)$-plane. 
All trial move parameters were set such that they each had an acceptance ratio 
of around 25-40\%.

Pressures were estimated across the mid-plane and were collected over 
$4\times 10^5$ Monte Carlo cycles for a given separation and $\Xi$. 
We applied block averaging of ten blocks to estimate the errors in pressures.
A Monte Carlo cycle consisted of either of 512 trial displacement of 
the point charges or a floppy-box move. 
Crystal structures were identified both by single configuration snapshots 
and ensemble averaged 2D-pair correlation maps (of the whole space, of each 
individual half-space, and across the half-spaces) in the $(x,y)$-plane.  

We found by numerical simulations that at finite coupling the particles 
form crystal of type I-III with the aspect-ratio parameter $\Delta$ which 
depends on both the coupling constant $\Xi$ and the interplate distance $d$, 
i.e., $\Delta(\Xi,d)$.
The lattice parameter $\Delta$ was obtained by finding the positions of 
the first two peaks of the 1D-pair correlation functions (ensured that they 
indeed form crystal structures) where only nearest $(x,y)$ neighbours,
identified by a Delaunay triangulation, been accounted for. 
The lattice parameter was then taken as the ratio between these two 
peak positions. 
Once the lattice parameter reaches unity it is not possible to extract it by 
this method as the two peaks coalesce to one peak, here we rely instead
of inspection of both the 2D- and 1D-pair correlation functions as well as 
single configuration snapshots to indeed verify that we had square structures 
(i.e., $\Delta=1$). 
Crystal structures could be identified for all studied $\eta$'s only for 
the largest $\Xi=175000$. 
$\Xi=17500$ only gave crystal structures for $\eta \ll 0.01$. 
This is consistent with previous numerical results \cite{Serr,Grimes79,Morf79} 
which predict 2D crystallization at any $d$ for coupling parameters above 
$\Xi\approx 31000$ and crystallization at contact ($d=0$) above 
$\Xi\approx 15600$.
The factor of two between both thresholds stems from the fact that at $d=0$, 
the two layers merge into one, with a double surface charge.

\renewcommand{\theequation}{4.\arabic{equation}}
\setcounter{equation}{0}

\section{Large-coupling description of the crystal phase} \label{Sec.crystal}
For a bilayer Wigner crystal, experiments \cite{Grimes79} and simulations 
\cite{Morf79} give the estimate $\Xi \simeq 31000$ for melting.  
This behavior follows from the restricted model in which counterions move 
only within the 2D Wigner single-layers. 
In this part, we shall consider $\Xi$ to be large enough to localize
particles near their Wigner-crystal positions. 
In our model, as soon as $\Xi$ is non divergent (finite $T$), 
the particles are not constrained to the wall surfaces
and can move in the whole slab domain $\Lambda$.
Within the canonical ensemble, the relevant thermodynamic quantities are 
the partition function $Z_N$ and the corresponding (dimensionless) 
free energy per particle $\beta f$ defined, up to some irrelevant constants
due to the background-charge density, as follows
\begin{equation} \label{partition}
Z_N = \frac{1}{N!} \int_{\Lambda} \prod_{i=1}^N \frac{{\rm d}^3r_i}{\lambda^3}\,
{\rm e}^{-\beta E(\{ {\bf r}_i\})} , \qquad
\beta f = - \frac{1}{N} \ln Z_N ,
\end{equation} 
where $E(\{ {\bf r}_i\})$ is the Coulomb interaction energy of the particles
and $\lambda$ stands for the thermal de Broglie wavelength.
We recall that the electric potential induced by the symmetrically charged 
plates is equal to 0 between the plates. 
The mean particle number density at point ${\bf r}$ is defined as 
$\rho({\bf r}) = \left\langle \sum_{i=1}^N 
\delta({\bf r}-{\bf r}_i) \right\rangle$,
where $\langle \cdots \rangle$ means the statistical average over 
the canonical ensemble.
It fulfills the sum rule $\int_{\Lambda} {\rm d}^3r\, \rho({\bf r}) = N$.
For our particle density which depends only on the perpendicular $z$-coordinate,
$\rho({\bf r}) \equiv \rho(z)$, this sum rule reduces to the electro-neutrality
condition $\int_0^d {\rm d}z\, \rho(z) = N/S = 2\sigma$.
The particle number density will be considered in a rescaled form
\begin{equation} \label{respart}
\widetilde{\rho}(\widetilde{z}) \equiv 
\frac{\rho(\mu\widetilde{z})}{2\pi\ell_{\rm B}\sigma^2} ,
\end{equation}
in terms of which the electro-neutrality condition  takes the form
\begin{equation} \label{normatilde}
\int_0^{\widetilde{d}} {\rm d}\widetilde{z}\, \widetilde{\rho}(\widetilde{z}) 
= 2 . 
\end{equation}

\subsection{Harmonic expansion} \label{Subsec.1}
The usual large-coupling approach to the counterion system between 
symmetrically charged plates is to make a harmonic expansion of particle 
coordinates around their Wigner bilayer positions \cite{Goldoni96}.
We found by numerical simulations that such an approach is not fully adequate
and one should assume that at non-infinite coupling, the particles form 
another reference crystal of type I-III with the aspect-ratio 
parameter $\Delta$ depending, besides the inter-plate distance $\eta$ as 
it was in the ground state, also on the coupling constant $\Xi$: 
$\Delta(\Xi,\eta)$.
In particular, the previously calculated infinite-coupling result  
$\Delta_0(\eta)$ in Refs. \cite{Goldoni96,Schweigert99,Messina03,Samaj12a} 
corresponds to $\Delta(\Xi\to\infty,\eta)$.
We aim at performing the harmonic expansion of particle coordinates around 
this reference crystal, evaluate the corresponding free energy and 
determine the $\Delta$-parameter of the reference crystal subsequently by
minimizing the free energy with respect to $\Delta$.
At finite coupling, the particles fluctuate around sites of the reference 
Wigner crystal, but as soon as the system is in its crystal phase,
the particle are localized close to these sites and the reference crystal 
is not an auxiliary theoretical construction, but its parameters are clearly 
visible in numerical experiments.     

Performing an expansion of the Coulombic energy up to quadratic order 
in particles displacements, we show in section \ref{Sssec:harmonic} of 
the Supplementary Information that
\begin{equation} \label{totalenergy}
E(\{ {\bf r}_i\}) = N e_0(\eta,\Delta) + \delta E , \qquad
\delta E = \sum_{i<j} \delta E_{ij} .
\end{equation}
with
\begin{equation} \label{deltaE}
- \beta \delta E = - \kappa(\eta,\Delta) 
\left[ \sum_{i\in\Sigma_1} \widetilde{z}_i + \sum_{i\in\Sigma_2} 
(\widetilde{d}-\widetilde{z}_i) \right] + \frac{1}{\sqrt{\Xi}} S_z
- \sqrt{\frac{\Xi}{2\pi}} \frac{\sigma}{2} \sum_{i<j}
\left[ B_{ij}^x (x_i-x_j)^2 + B_{ij}^y (y_i-y_j)^2 \right] + \cdots . 
\end{equation}
Here, the prefactor of the linear terms in $\widetilde{z}_i$ or 
$(\widetilde{d}-\widetilde{z}_i)$ reads
\begin{eqnarray}
\kappa(\eta,\Delta) & = & \frac{\eta}{2\pi} \sum_{i_x,i_y} 
\frac{\Delta^{3/2}}{\left[ (i_x-1/2)^2 + \Delta^2 (i_y-1/2)^2
+\Delta\eta^2\right]^{3/2}} \nonumber \\
& = & \frac{\eta}{\pi^{3/2}} \int_0^{\infty} {\rm d}t\, \sqrt{t}\, 
{\rm e}^{-\eta^2 t} \theta_2\left({\rm e}^{-t\Delta}\right) 
\theta_2\left({\rm e}^{-t/\Delta}\right) = 
- \frac{1}{2 \pi^{3/2}} \frac{\partial}{\partial\eta} \Sigma(\eta,\Delta) + 1 .
\label{kappa}
\end{eqnarray}
The quantities $S_z$ and $B_{ij}$ are given by Eqs. \eqref{Sz}, \eqref{Bsame} 
and \eqref{Bdifferent}, while the Jacobi theta function $\theta_2$ is defined as
$\theta_2(q) = \sum_j q^{\left(j-\frac{1}{2}\right)^2}$ (see the Supplementary
Information where $\theta_3(q) = \sum_j q^{j^2}$ is also required).

The particle coordinates $\{ x_i\}$, $\{ y_i\}$ and $\{ z_i\}$ are decoupled 
within the harmonic expansion of the energy change (\ref{deltaE}).
Within the present formalism, the particles have a well defined 
appurtenance to plate $\Sigma_1$ or $\Sigma_2$ in the Wigner bilayer.
The leading term in the $z$-subspace is linear in $\widetilde{z}_i$ 
for particles $i\in \Sigma_1$ and in $(\widetilde{d}-\widetilde{z}_i)$ 
for particles $i\in \Sigma_2$, with the prefactor function $\kappa$ depending 
on $\eta$ and $\Delta$.
This effective electric one-body potential subsumes the effects of the uniform 
surface charges on the two plates and the particle layer on the opposite plate, 
while particles on the same plate contribute to higher-order quadratic terms.
In the limit of small inter-plate distance $\eta\to 0$, we have 
\begin{equation}
\lim_{\eta\to 0} \kappa(\eta,\Delta) = 0 ,
\end{equation}
i.e. each particle feels the zero potential coming from the uniform surface 
charge densities on the two plates while the effect of the opposite particle 
layer with the lattice spacing $a\gg d$ is negligible; this description 
coincides with the standard one-body SC fugacity approach for two 
symmetrically charged plates at small distances
\cite{Moreira00,Netz01,Moreira02}. 
In the large distance limit $\eta\to\infty$ we have 
\begin{equation}
\lim_{\eta\to \infty} \kappa(\eta,\Delta) = 1 ,
\end{equation}
i.e. each particle feels the linear electrostatic potential, $\widetilde{z}$ 
or $(\widetilde{d}-\widetilde{z})$, coming from the surface charge at 
its own plate; at large distances the discrete counterion structure on 
the opposite plate is seen as a charge continuum neutralized by 
the opposite background charge on that plate.
In this way the $\kappa$-function describes correctly a continuous 
interpolation from a two-plate picture at $\eta\to 0$ to a one-plate picture 
at $\eta\to\infty$.
The contribution of quadratic terms in $S_z/\sqrt{\Xi}$, which becomes 
negligible in comparison with the one-body ones in the SC limit $\Xi\to\infty$,
will be treated perturbatively for large $\Xi$.
The quadratic terms in the $(x,y)$-plane reflect strong particle 
correlations/repulsions in this plane.
Due to the strong particle repulsions, it is reasonable to constrain 
the particle coordinates within one elementary cell, i.e. 
\begin{equation} \label{cell}
- \frac{a}{2} < x_i < \frac{a}{2} , \qquad 
- \frac{a\Delta}{2} < y_i < \frac{a\Delta}{2} .
\end{equation}

The partition function (\ref{partition}), with the particle interaction energy 
given by Eqs. (\ref{totalenergy}) and (\ref{deltaE}), factorizes into
\begin{equation}
Z_N = \frac{1}{N!} \left( \frac{\mu}{\lambda} \right)^N 
\exp\left[ -\beta N e_0 \right] Q_z Q_x Q_y ,
\end{equation}
where 
\begin{eqnarray}
Q_z(\eta,\Delta) & = & \int_0^{\widetilde{d}} \prod_{i\in\Sigma_1} 
{\rm d}\widetilde{z}_i\, {\rm e}^{-\kappa \widetilde{z}_i} \int_0^{\widetilde{d}} 
\prod_{i\in\Sigma_2} {\rm d}\widetilde{z}_i\,
{\rm e}^{-\kappa(\widetilde{d}-\widetilde{z}_i)} \exp(S_z) , \label{Qz} \\
Q_x(\eta,\Delta) & = & \int_{-a/2}^{a/2} \prod_{i\in\Sigma_1\cup \Sigma_2} 
\frac{{\rm d}x_i}{\lambda}\, \exp\left[ - \sqrt{\frac{\Xi}{2\pi}} 
\frac{\sigma}{2} \sum_{i<j} B_{ij}^x (x_i-x_j)^2 \right] , \label{Qx} \\
Q_y(\eta,\Delta) & = & \int_{-\Delta a/2}^{\Delta a/2} \prod_{i\in\Sigma_1\cup \Sigma_2} 
\frac{{\rm d}y_i}{\lambda}\, \exp\left[ - \sqrt{\frac{\Xi}{2\pi}} 
\frac{\sigma}{2} \sum_{i<j} B_{ij}^y (y_i-y_j)^2 \right] . \label{Qy}
\end{eqnarray} 
From now on we shall automatically neglect irrelevant terms which do not 
depend on $\eta$ and $\Delta$.
The free energy per particle is given in the harmonic approximation by
\begin{equation} \label{freerepr}
\beta f(\eta,\Delta) = \frac{\sqrt{\Xi}}{2^{3/2}\pi} \Sigma(\eta,\Delta) - 
\frac{1}{N} \ln Q_z - \frac{1}{N} \ln Q_x - \frac{1}{N} \ln Q_y .
\end{equation}

\subsection{Thermodynamics} \label{Subsec.2}
Obtaining the partial partition functions $Q_x$, $Q_y$ and $Q_z$ is 
a non-trivial task, performed in section \ref{Sssec:thermo} in 
the Supplementary Information.
It relies on the diagonalization of the inverse variance-covariance matrices
of fluctuations in the $x$, $y$, and $z$ coordinates, which is achieved 
by means of a 2D Fourier transform.
The resulting free energy per particle $f$ is expressible 
in the harmonic approximation as
\begin{equation}
\beta f(\eta,\Delta) = \beta f^{(0)}(\eta,\Delta) + \frac{1}{\sqrt{\Xi}}
\beta f^{(1)}(\eta,\Delta) + O\left( \frac{1}{\Xi}\right) .
\end{equation}
The leading WSC term reads as
\begin{eqnarray}
\beta f^{(0)}(\eta,\Delta) & = & \frac{\sqrt{\Xi}}{2^{3/2}\pi} 
\Sigma(\eta,\Delta) -
\ln \left( \frac{1-{\rm e}^{-\kappa(\eta,\Delta)\widetilde{d}}}{\kappa} \right)
\nonumber \\ & & + \frac{1}{4} \int_0^1 {\rm d}q_x \int_0^2 {\rm d}q_y\,
\ln\left\{ \left[ C^x(0,0) - C^x(q_x,q_y-q_x) \right] 
\left[ C^y(0,0) - C^y(q_x,q_y-q_x) \right] \right\} , \label{freeenergy}
\end{eqnarray}
where the functions $ C^x({\bf q})$ and $ C^y({\bf q})$ are given by Eqs.
\eqref{eq:cx} and \eqref{eq:cy}.
The prefactor function to the first correction
$\beta f^{(1)}(\eta,\Delta) = - \langle S_z \rangle_0/N$
is given by Eqs. (\ref{Sz0})-(\ref{F}).
All quantities in the above formulas are expressed as fast converging 
series of generalized Misra functions.
This means that the thermodynamics can be treated on the same footing as 
the ground-state energy, at least in the harmonic approximation.

According to the principle of minimum free energy, the aspect ratio
of the rectangular lattice $\Delta$ is fixed by the condition
\begin{equation} \label{varDelta}
\frac{\partial}{\partial \Delta} \beta f(\eta,\Delta) = 0 
\end{equation}
which provides the explicit dependence of $\Delta$ on the coupling constant 
$\Xi$ and the plate distance $\eta$, $\Delta(\Xi,\eta)$.
Compare this relation with its ground-state counterpart Eq. (\ref{Delta0})
which reflects an analogous minimization of the interaction energy.

The pressure exerted on the plates can be obtained via
the thermodynamic route as follows
\begin{equation}
\beta P_{\rm th} = - \frac{\partial}{\partial d} 
\left( \frac{\beta F}{S} \right)
= - 2 \sigma^{3/2} \frac{\partial (\beta f)}{\partial\eta} .
\end{equation}
Rescaling the pressure in the same way as the particle density in 
(\ref{respart}), we get
\begin{equation} \label{Pth}
\widetilde{P}_{\rm th} \equiv \frac{\beta P_{\rm th}}{2\pi\ell_{\rm B}\sigma^2}
= - \sqrt{\frac{2}{\pi\Xi}} \frac{\partial}{\partial\eta} 
\left[ \beta f(\eta,\Delta) \right] .
\end{equation}
The positive/negative sign of the pressure means an effective 
repulsion/attraction between the charged walls.

\subsection{Particle density profile and pressure} \label{Subsec.3}
To find the particle density, we add to each particle in the Hamiltonian 
the generating (source) one-body potential $u({\bf r})$ which will be set 
to 0 at the end of calculations.
The partition function (\ref{partition}) is then transformed to
\begin{equation} \label{partitiongen}
Z_N[w] = \frac{1}{N!} \int_{\Lambda} \prod_{i=1}^N \frac{{\rm d}{\bf r}_i}{
\lambda^3}\, w({\bf r}_i) {\rm e}^{-\beta E(\{ {\bf r}_i\})}
\end{equation} 
and it is a functional of the generating Boltzmann weight
$w({\bf r}) = \exp[-\beta u({\bf r})]$.
The particle density at point ${\bf r}$ is then obtained as the functional
derivative:
\begin{equation}
\rho({\bf r}) = \frac{\delta}{\delta w({\bf r})} \ln Z_N[w] 
\Big\vert_{w({\bf r})=1} .
\end{equation}
We show in the Supplementary Information that the (rescaled) particle density 
takes the WSC expansion form
\begin{equation} \label{densitySC}
\widetilde{\rho}(\widetilde{z}) = \widetilde{\rho}^{(0)}(\widetilde{z}) 
+ \frac{1}{\sqrt{\Xi}} \widetilde{\rho}^{(1)}(\widetilde{z}) + \cdots ,
\end{equation}
with the leading WSC order
\begin{equation} \label{rho0}
\widetilde{\rho}^{(0)}(\widetilde{z}) = 
\frac{\kappa}{1-{\rm e}^{-\kappa\widetilde{d}}} \left[ {\rm e}^{-\kappa\widetilde{z}} 
+ {\rm e}^{-\kappa(\widetilde{d}-\widetilde{z})} \right] .
\end{equation} 
This leading WSC particle density has the correct reflection 
$\widetilde{z}\to (\widetilde{d}-\widetilde{z})$ symmetry and satisfies 
the expected normalization condition
\begin{equation} \label{normatildeSC}
\int_0^{\widetilde{d}} {\rm d}\widetilde{z}\, 
\widetilde{\rho}^{(0)}(\widetilde{z}) = 2 . 
\end{equation}
The first correction to the particle density is given in Eq. \eqref{eq:rho1}.
Note that $\int_0^{\widetilde{d}} {\rm d}\widetilde{z}\, 
\widetilde{\rho}^{(1)}(\widetilde{z}) = 0.$
The same property holds also in higher WSC orders, so that 
the electroneutrality condition (\ref{normatilde}) is ensured on
the leading WSC order (\ref{normatildeSC}).

Invoking the contact theorem for planar walls \cite{contact,MaTT15}, 
we obtain the pressure as
\begin{equation} \label{ct}
\widetilde{P}_{\rm c} = \widetilde{\rho}(0) - 1 = \left[ 
\widetilde{\rho}^{(0)}(0)-1 \right] + \frac{1}{\sqrt{\Xi}} 
\widetilde{\rho}^{(1)}(0) + \cdots .
\end{equation}
Writing the WSC expansion for the ``contact'' pressure as
$\widetilde{P}_{\rm c} = \widetilde{P}_{\rm c}^{(0)} + 
\widetilde{P}_{\rm c}^{(1)} / \sqrt{\Xi} + \cdots$,
we have in the leading order
\begin{equation} \label{P0}
\widetilde{P}_c^{(0)} = \kappa \left( \frac{1+{\rm e}^{-\kappa\widetilde{d}}}{
1-{\rm e}^{-\kappa\widetilde{d}}} \right) - 1 ,
\end{equation}
The first correction is given in section \ref{Sssec:density} of 
the Supplementary Information.

Since $\kappa\to 1$ for $\widetilde{d}\to \infty$ it is simple to show
that the expansion coefficients $\widetilde{P}_c^{(0)}$ and 
$\widetilde{P}_c^{(1)}$ vanish in the asymptotic large-distance limit, 
as they should.
The thermodynamic $\widetilde{P}_{\rm th}$ and contact $\widetilde{P}_{\rm c}$ 
pressures must coincide in an exact theory.
In an approximate theory like ours, the difference between the two pressures
indicates the accuracy of the approach. 

\subsection{Comparison with numerical results} \label{Subsec.4}
\begin{figure}[tbp]
\begin{center}
\includegraphics[width=0.45\textwidth,clip]{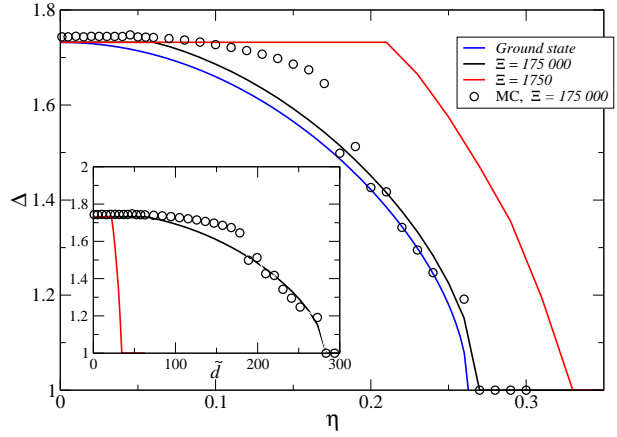}
\caption{Dependence on the dimensionless distance between the walls $\eta$ 
of the aspect ratio $\Delta$ for structures I-III. 
The MC data for the coupling constant $\Xi=175000$ are indicated by open 
circles. 
The results of the present WSC theory for $\Xi=175000$ and $\Xi=1750$ are 
represented by black and red curves, respectively. 
The ground-state plot, blue curve, is also given as a reference.
For comparison, the inset shows the plots of $\Delta$ versus 
$\widetilde{d}$ for $\Xi=175000$ (black curve) and $\Xi=1750$ (red curve).}
\label{fig:Deltavseta} 
\end{center}
\end{figure}

We compare the results of our WSC theory with MC data for two
values of the coupling constant, namely for large $\Xi=175000$ when the system 
is in its crystal phase and small $\Xi=1750$ when the system behaves as a fluid.
The distance dependence of the aspect ratio of the rectangular lattice $\Delta$ 
is pictured in Fig. \ref{fig:Deltavseta}.
The ground-state case ($\Xi\to\infty$) is represented by the blue curve.
The results of the WSC theory are shown by the black curve for $\Xi=175000$ 
and by the red curve for $\Xi=1750$; note that on the scale of our graph 
the results of the leading order and the leading order plus the first 
correction are indistinguishable.
In contrast to the ground state with phase I ($\Delta=\sqrt{3}$) occurring 
only at $\eta=0$ \cite{Messina03,Samaj12a}, phase I exists in a finite
interval of $\eta$: up to $\eta\approx 0.06$ for $\Xi=175000$ and
up to $\eta\approx 0.21$ for $\Xi=1750$. 
The second-order phase transition between phases I and II is of mean-field
type, with $\sqrt{3}-\Delta$ the order parameter.
The MC data for $\Xi=175000$ are represented by open circles. 
They agree qualitatively with our theoretical results, namely phase I
is dominant up to $\eta\approx 0.07$.
For the smaller coupling constant $\Xi=1750$, the $\eta$-range where 
the structures I-II (and also III) prevail increases; in MC simulations,
we did not identify any crystal phase and the counterion system behaves as 
a fluid.  
In the inset of Fig. \ref{fig:Deltavseta}, we plot the two theoretical curves 
and MC data for $\Delta$ versus $\widetilde{d}$; we see that the two 
theoretical curves differ much from one another in this representation.
We recall here that the connexion between both scales 
reads $\widetilde d = \eta\,\sqrt{2\pi\Xi}$.

\begin{figure}[tbp]
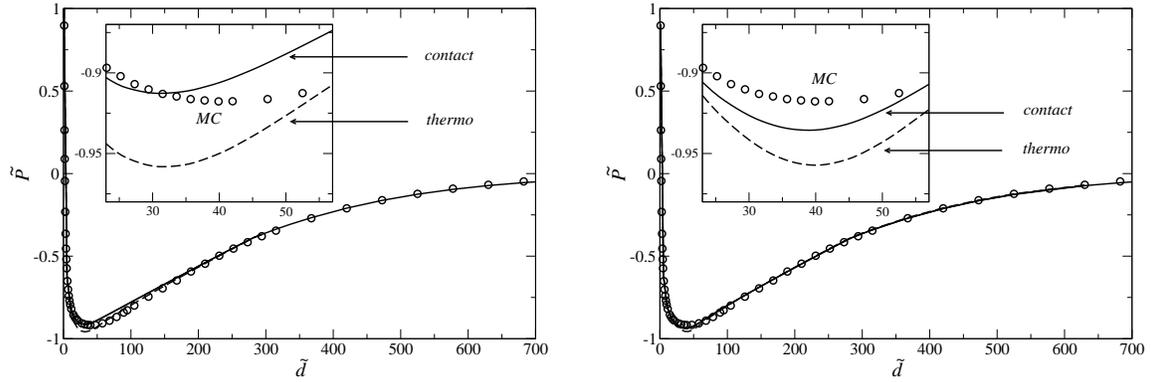

\begin{center}
\includegraphics[width=0.4\textwidth,clip]{P_Xi175000_nocorr.eps}
\qquad
\includegraphics[width=0.4\textwidth,clip]{P_Xi175000_corr.eps}
\caption{Dependence of the (rescaled) pressure $\widetilde{P}$
on the dimensionless distance $\widetilde{d}$ at
$\Xi=175000$.
The left panel corresponds to the leading WSC order, the right panel
to the leading WSC order plus the first correction.
The MC data are indicated by open circles. 
The pressures obtained via the thermodynamic route and by using the contact 
theorem are represented by dashed and solid curves, respectively.
The insets magnify the regions around the pressure minimum.}
\label{fig:P_Xi175000} 
\end{center}
\end{figure}

\begin{figure}[tbp]
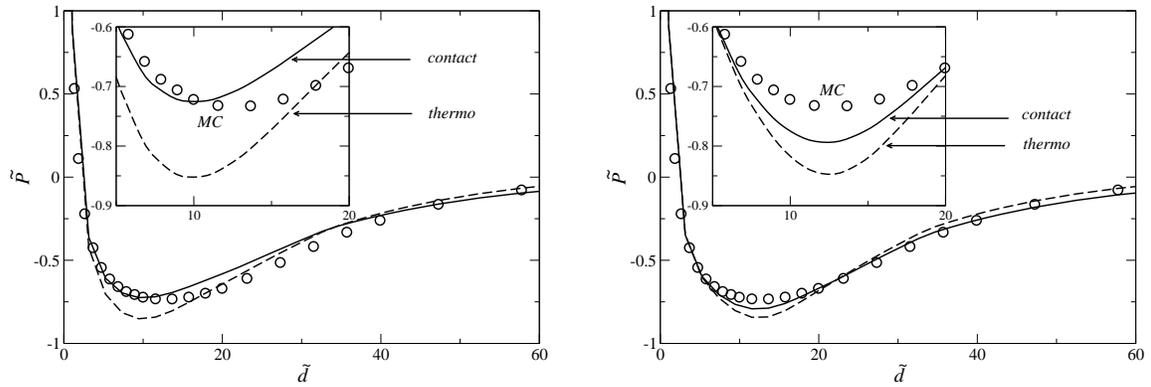

\begin{center}
\includegraphics[width=0.4\textwidth,clip]{P_Xi1750_nocorr.eps}
\qquad
\includegraphics[width=0.4\textwidth,clip]{P_Xi1750_corr.eps}
\caption{Same as Fig. \ref{fig:P_Xi175000} for $\Xi=1750$.}
\label{fig:P_Xi1750} 
\end{center}
\end{figure}

The distance dependence of the  pressure  is presented in 
Fig. \ref{fig:P_Xi175000} for the coupling constant $\Xi=175000$.
The left panel corresponds to the leading WSC order, the right panel
presents the results of the leading WSC order plus the first correction.
The MC data are indicated by open circles. 
The WSC results obtained by the thermodynamic route and by the contact 
theorem are represented by the dashed and solid curves, respectively.
It is seen that data obtained by the two methods are very close 
to one another, and to the MC measures. 
The location and the value of the pressure minimum is determined
especially well by the WSC theory including the first SC correction
(see the insets). 
A very good coincidence with the MC data lasts up to extremely large
values of $\widetilde{d}$, corresponding to $\eta\approx 1$,
well beyond the validity of the standard fugacity
\cite{Moreira00,Netz01,Moreira02} and Wigner-crystal \cite{Samaj11}
SC approaches.   
The analogous plots of $\widetilde{P}$ versus $\widetilde{d}$ for 
the intermediate value of the coupling constant $\Xi=1750$ are presented in 
Fig. \ref{fig:P_Xi1750}.
In spite of the fact that the counterion system is in the fluid state for this 
value of $\Xi$, the analytic results agree surprisingly well with MC data.
A similar conclusion holds at even smaller $\Xi$ values, 
see section \ref{Sec.conclusion} where we present data at $\Xi=50$. 
This points to the fact that what is relevant is not so much the detailed 
ionic configuration, but that it is strongly modulated. 
This gives support to the idea of a correlation hole, 
developed in section \ref{Sec.fluid}.

\begin{figure}[tbp]
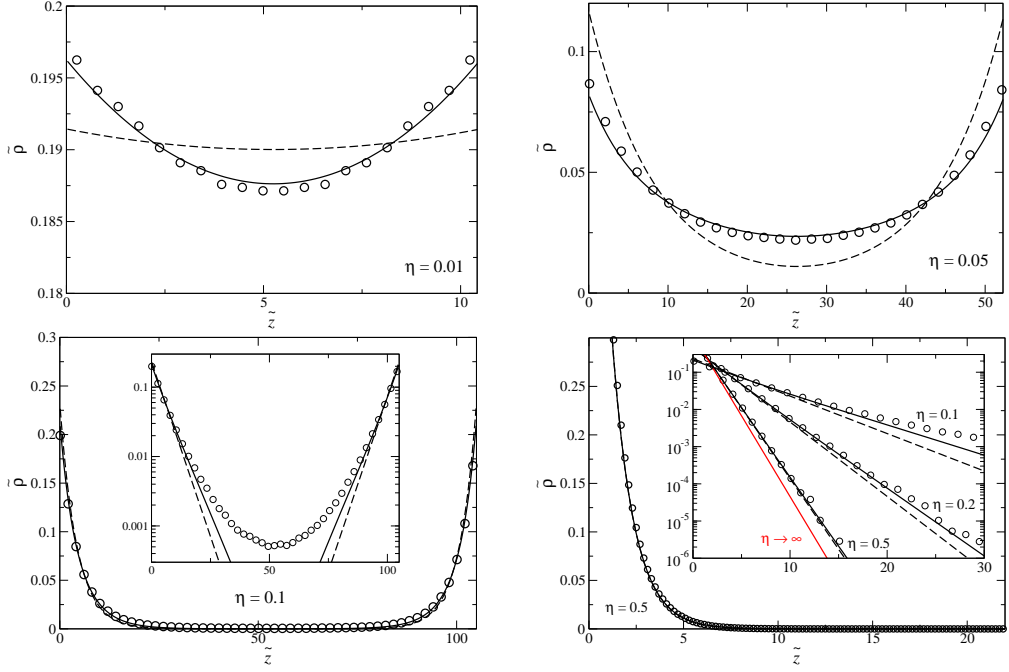

\begin{center}
\includegraphics[width=0.35\textwidth,clip]{rho_z_Xi175000_eta0.01.eps} \qquad
\includegraphics[width=0.35\textwidth,clip]{rho_z_Xi175000_eta0.05.eps} \qquad 
\includegraphics[width=0.35\textwidth,clip]{rho_z_Xi175000_eta0.1.eps}  \qquad
\includegraphics[width=0.35\textwidth,clip]{rho_z_Xi175000_eta0.5.eps}
\caption{Rescaled particle density $\widetilde{\rho}$ versus the dimensionless
coordinate $\widetilde{z}$ for $\Xi=175000$ and the 
(dimensionless) distance between the walls $\eta=0.01$, $0.05$, $0.1$ and $0.5$.
As before, the MC data are indicated by open circles.
The density profiles obtained in the leading WSC order and in 
the leading WSC order plus the first correction are represented by 
the dashed and solid curves, respectively.
The logarithmic plots in the insets document the slopes of the density
profile close to the walls.
The one-wall density profile ($\eta\to\infty$, red line) is pictured 
for illustration.}
\label{fig:rho_z_Xi175000} 
\end{center}
\end{figure}

At $\Xi=175000$, some representative particle density profiles are pictured in
Fig. \ref{fig:rho_z_Xi175000}. 
The MC data are indicated by open circles.
The density profiles obtained in the leading WSC order, see Eq. (\ref{rho0}), 
and with the first correction included, see Eqs. (\ref{densitySC}) and 
(\ref{rho1}), are represented by the dashed and solid curves, respectively.
The logarithmic plots in the insets illustrate that in the large
interval of $\eta=0.1-0.5$ the present WSC theory predicts contact particle 
densities and the slopes of the density profile close to the walls which are 
in excellent agreement with MC data. 
This shows the relevance of the idea of an effective local field
(an effective one-body potential), embodied in $\kappa$, which depends on 
the distance and on the geometry of the ionic arrangement, together with 
the accuracy of our approach for computing this non-trivial quantity.
Besides, it is noteworthy that for the considered extremely large coupling 
constant, the inclusion of the correction to the leading WSC order improves 
substantially the results. 
For small distances $\eta=0.01$ and $0.05$, the WSC density profiles
agree with MC in the whole inter-plate slab, but with increasing 
$\eta$ there is a discrepancy between the WSC and MC results in the middle 
region between the walls characterized by extremely small particle densities.

\renewcommand{\theequation}{5.\arabic{equation}}
\setcounter{equation}{0}

\section{Strong-coupling theory for the fluid phase} 
\label{Sec.fluid}
The Wigner bilayer is stable at very large values of the coupling constant.
For intermediate and small values of $\Xi$, the counterion 
system behaves as a fluid which is isotropic along the $(x,y)$ plane.
The strong Coulomb repulsion leads to a depletion region around each particle,
inaccessible to other particles, known as the correlation hole
\cite{Chen06,Rouzina96,Nordholm84,Santangelo06,Hatlo10,Samaj16}.

Within the WSC theory, the Wigner structure in the $(x,y)$ plane underlies
the calculation of the crucial effective local field $\kappa(\eta,\Delta)$,
see Eq. (\ref{kappa}). 
It determines the slope of the density profile close to the wall. 
To describe physically the fluid regime, the idea is to substitute the lattice 
representation of $\kappa$ by its continuum counterpart, with a radial cut 
of the lattice summation at small distances $R$ due to the correlation hole.
In particular, rewriting the lattice sum as
\begin{equation} \label{rewrite}
\sum_{i_x,i_y} \frac{\Delta^{3/2}}{\left[ (i_x-1/2)^2 + \Delta^2 (i_y-1/2)^2
+\Delta\eta^2\right]^{3/2}} = \frac{1}{\sigma^{3/2}}
\sum_{i_x,i_y} \frac{1}{\left[ a^2(i_x-1/2)^2 + \Delta^2 a^2(i_y-1/2)^2
+\Delta a^2 \eta^2\right]^{3/2}}   
\end{equation}
and regarding that there is surface $a^2\Delta=1/\sigma$ per site on 
the Wigner lattice, we can express (\ref{rewrite}) as a continuum integral 
in the following way
\begin{equation}
\frac{1}{\sigma^{3/2}} \frac{1}{1/\sigma} \int_R^{\infty} {\rm d}r\, 2\pi r
\frac{1}{\left( r^2 + \frac{\eta^2}{\sigma}\right)^{3/2}}
= \frac{2\pi}{\sqrt{\eta^2+\sigma R^2}} . 
\end{equation}
To estimate the short-distance cut $R$, i.e. the radius of the correlation 
hole around the reference particle, one has to realize that the reference 
particle on plate 1 is in the center of an elementary cell of the particle 
crystal on plate 2.
Let us choose the symmetric $\Delta=1$ square lattice, and apply the Voronoi 
construction of the Wigner-Seitz primitive cell which has  surface 
$a^2/2=1/(2\sigma)$. 
Thus, $\pi R^2 = 1/(2\sigma)$ and we end up with 
\begin{equation} \label{kappafluid1}
\kappa(\eta) = \frac{\eta}{\sqrt{\eta^2 + \frac{1}{2\pi}}} .
\end{equation}
This fluid version of the $\kappa$-function has the correct limiting
values $\kappa=0$ for $\eta\to 0$ and $\kappa=1$ for $\eta\to\infty$.
We shall refer to this correlation-hole theory to as ch1.

Another phenomenological way to express the functional dependence of
$\kappa(\eta)$ combines geometrical features, overall electroneutrality 
together with space fluctuations of charged particles in the fluid regime.
We substitute the crystal bilayer structure by a couple of correlation holes 
with respect to a reference particle, which appertains say to plate 1. 
We thereby obtain one disk of radius $R_1$ at plate 1 and the other disk of 
radius $R_2$ at plate 2. 
Particles are smeared out on the plate regions outside of the correlation-hole
disks; the corresponding ``punctuated'' planes are therefore taken as neutral.
The charge of the reference particle must be compensated by the total
surface charge on the disks which implies the constraint
\begin{equation} \label{chrule}
-e + \sigma e \left( \pi R_1^2 + \pi R_2^2 \right) = 0 .  
\end{equation}
The disk radiuses depend on the distance between the plates, $R_1=R_1(\eta)$
and $R_2=R_2(\eta)$.
If the two walls touch each other, $\eta=0$, the correlation holes 
around the reference particle are the same on both sides, i.e.,
\begin{equation}
R_1^2(\eta=0) = R_2^2(\eta=0) = \frac{1}{2\pi\sigma} .
\end{equation}
The plate-1 and plate-2 subspaces decouple at asymptotically large distances 
$\eta\to\infty$. 
From the point of view of the reference particle (attached to plate 1), 
the hole at plate 2 disappears due to thermal fluctuations of charged particles 
at plate 2, $R_2(\eta\to\infty)=0$, while the charge conservation rule
$-e + \sigma e \pi R_1^2(\eta\to\infty) = 0$ leads to an increase of 
the radius of the hole at plate 1: $R_1^2(\eta\to\infty)=1/(\pi\sigma)$,
like in the one-plate geometry.
Respecting the constraint (\ref{chrule}), the two limits are matched by 
the phenomenological interpolation formulas
\begin{equation}
R_1^2(\eta) = \frac{1}{2\pi\sigma} + \frac{1}{2\pi\sigma} 
\frac{\eta}{c+\eta} , \qquad
R_2^2(\eta) = \frac{1}{2\pi\sigma} - \frac{1}{2\pi\sigma} 
\frac{\eta}{c+\eta} ,
\end{equation}
where $c$ defines a crossover scale. 
For simplicity we set $c=1$.
For the reference particle at distance $z$ from plate 1 and at distance 
$(d-z)$ from plate 2, the electrostatic energy $E(z)$ yielded by 
the two correlation holes is given by
\begin{eqnarray}
-\beta E(z) & = & \sigma \ell_{\rm B} \left[
\int_0^{R_1} {\rm d}r\, 2\pi r \frac{1}{\sqrt{r^2+z^2}} +
\int_0^{R_2} {\rm d}r\, 2\pi r \frac{1}{\sqrt{r^2+(d-z)^2}} \right] \nonumber \\
& = & \frac{1}{\mu} \left[ \sqrt{R_1^2+z^2} + \sqrt{R_2^2+(d-z)^2} - d \right] .
\end{eqnarray}
Within the single-particle picture, we can take the whole one-body Boltzmann 
factor $\exp[-\beta E(z)]$ or restrict ourselves to the linear term 
in the energy, $\exp(-\kappa\widetilde z)$, with
\begin{equation} \label{kappafluid2}
\kappa(\eta) = \frac{\eta}{\sqrt{\eta^2+\sigma R_2^2}}
= \frac{\eta}{\sqrt{\eta^2 + \frac{1}{2\pi(1+\eta)}}} .
\end{equation}
This $\kappa$ coincides with the geometrical one (\ref{kappafluid1}) at small 
distances $\eta\to 0$; it furthermore shares with ch1 the correct limiting 
value 1 at $\eta\to\infty$.
We shall refer to this correlation-hole theory to as ch2.

\begin{figure}[tbp]
\begin{center}
\includegraphics[width=0.4\textwidth,clip]{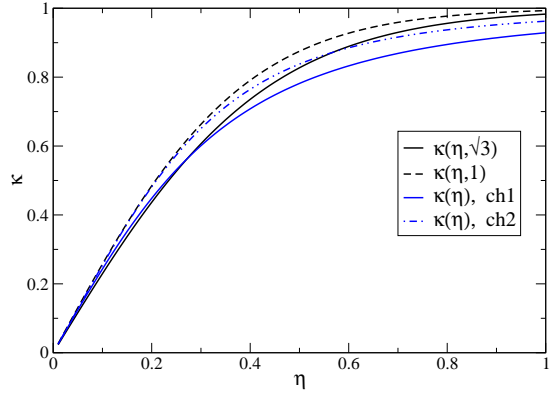}
\caption{Dependence of the effective local field $\kappa$, on rescaled distance.
The black solid and dashed curves correspond to two extreme crystal versions
with $\Delta=\sqrt{3}$ and $\Delta=1$, respectively.
The blue solid and dotted-dashed curves are for the correlation-hole
ch1 and ch2 theories, respectively.}
\label{fig:kappa} 
\end{center}
\end{figure}

The dependences of different variants of the function $\kappa$ on $\eta$  
are pictured in Fig. \ref{fig:kappa}.
The crystal versions of $\kappa(\eta,\Delta)$ with the extreme values
of the aspect ratio $\Delta=\sqrt{3}$ and $\Delta=1$ are represented by
the black solid and dashed curves, respectively. 
The blue solid and dotted-dashed curves correspond to the correlation-hole
ch1 formula (\ref{kappafluid1}) and the ch2 formula (\ref{kappafluid2}),
respectively.
Note that the four plots are relatively close to each other, which
documents the robustness of the method.   

\begin{figure}[tbp]
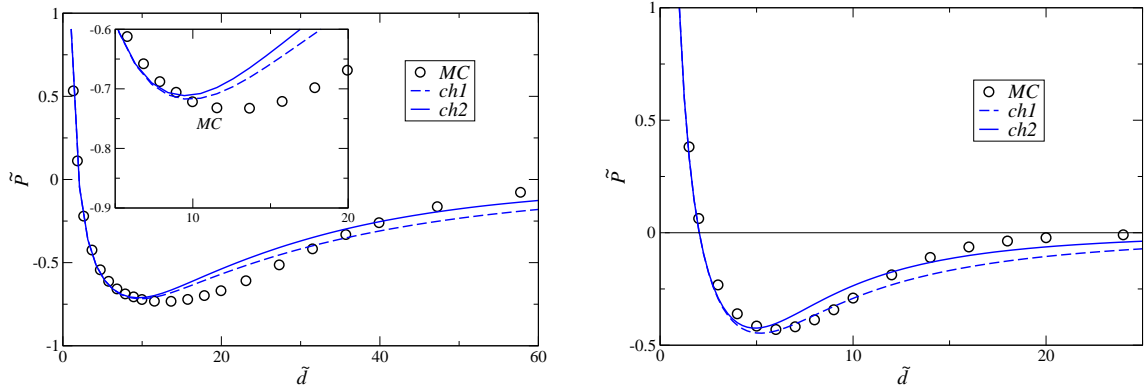

\begin{center}
\includegraphics[width=0.4\textwidth,clip]{P_Xi1750_ch.eps} \qquad
\includegraphics[width=0.4\textwidth,clip]{P_Xi100.eps}
\caption{Rescaled pressure $\widetilde{P}$ versus the dimensionless distance 
$\widetilde{d}$ for $\Xi=1750$ (left panel) and $\Xi=100$ (right panel).
The blue dashed and solid curves correspond to the correlation-hole
ch1 and ch2 theories, respectively.
The inset in the left panel magnifies the region around the pressure minimum.}
\label{fig:P_ch} 
\end{center}
\end{figure}

Having an expression for the fluid $\kappa(\eta)$, the leading SC estimate 
for the density profile is given by Eq. (\ref{rho0}) and the pressure can be 
obtained by using the contact formula (\ref{P0}).
For an intermediate coupling constant $\Xi=1750$, the plot of
the rescaled pressure $\widetilde{P}$ on $\widetilde{d}$ is pictured in
the left panel of Fig. \ref{fig:P_ch}. 
We see that the results of our two correlation-hole approaches
ch1 and ch2, represented respectively by the dashed and solid curves, are 
close to the MC data (open circle symbols). 
For the relatively small value of the coupling constant $\Xi=100$, 
the analogous plot is presented in the right panel, with again a fair agreement.

\begin{figure}[tbp]
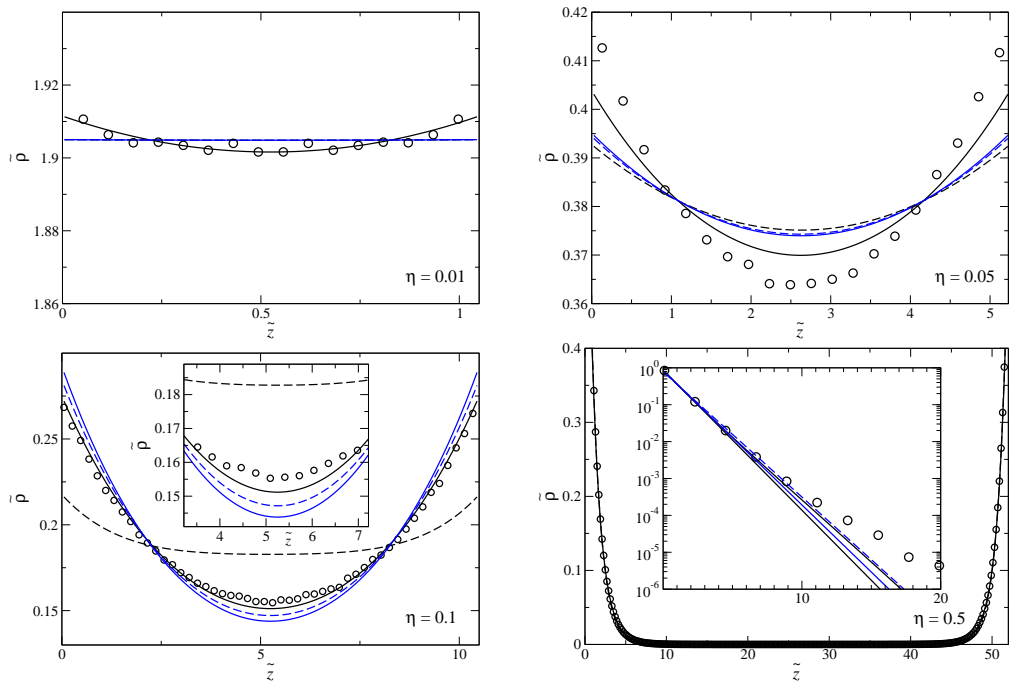

\begin{center}
\includegraphics[width=0.35\textwidth,clip]{rho_z_Xi1750_eta0.01.eps} \qquad
\includegraphics[width=0.35\textwidth,clip]{rho_z_Xi1750_eta0.05.eps} \qquad 
\includegraphics[width=0.35\textwidth,clip]{rho_z_Xi1750_eta0.1.eps}  \qquad
\includegraphics[width=0.35\textwidth,clip]{rho_z_Xi1750_eta0.5.eps}
\caption{Rescaled particle density $\widetilde{\rho}$ versus the dimensionless
coordinate $\widetilde{z}$ for $\Xi=1750$, at the indicated four 
distances between the walls. 
The MC data are shown by open circles.
The density profiles obtained in the leading WSC order and in 
the leading WSC order plus the first correction are represented by 
the black dashed and solid curves, respectively.
The blue dashed/solid curves correspond to the correlation-hole
ch1/ch2 theories.}
\label{fig:rho_z_Xi1750} 
\end{center}
\end{figure}

The density profiles for $\Xi=1750$ at the four
distances between the walls $\eta=0.01$, $\eta=0.05$, $\eta=0.1$ and 
$\eta=0.5$ are pictured in Fig. \ref{fig:rho_z_Xi1750}.
MC results are compared to the WSC predictions (with and without 
the first correction) as well as to the two correlation hole theories.
A conclusion that emerges is that while all approaches proposed yield 
acceptable quantitative results, the Wigner SC method is the most accurate.
This is somewhat surprising since we sit here in a coupling-range where 
no crystal is formed. Yet, accuracy requires that the correction is included,
and it stems from a rather demanding analytical work. Here, a fair 
assessment of ch performance would be to compare to WSC without 
correction, in which case ch is quite superior.

\renewcommand{\theequation}{6.\arabic{equation}}
\setcounter{equation}{0}

\section{Conclusion} \label{Sec.conclusion}
The aim of this paper was to construct a strong-coupling theory for thermal 
equilibrium of pointlike counterions between parallel and symmetrically charged 
plates. 
The goal was to extend significantly the range of interplate distances $d$
where a trustworthy analytical effective force is available. 
This range was hitherto reduced to $d$ smaller than a couple of Gouy-Chapman 
lengths \cite{Moreira02,Samaj11}, meaning $\widetilde d$ of order unity. 
To this end, we studied the counterion system in both the crystal phase 
at extremely large Coulombic couplings and in the fluid phase, at large and 
intermediate couplings.

\begin{figure}[tbp]
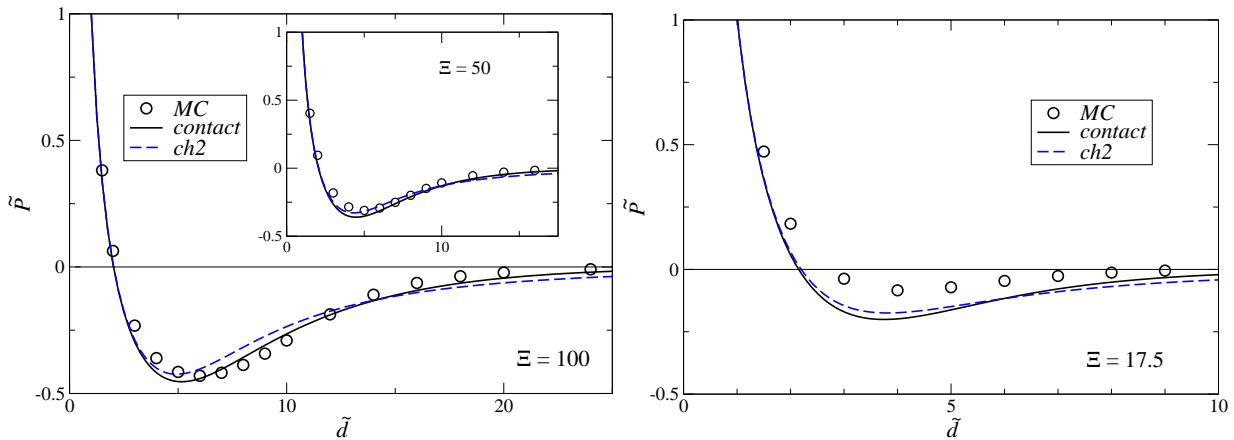

\begin{center}
\includegraphics[width=0.45\textwidth,clip]{P_Xi100final.eps} 
\includegraphics[width=0.45\textwidth,clip]{P_Xi17.5.eps} 
\caption{Testing the relevance of the WSC and correlation-hole calculations at $\Xi=50$ and $\Xi=100$.
The leading ``contact'' pressure computed from the Wigner SC approach 
is compared to Monte Carlo data and to the ch2 theory. The good agreement observed 
here deteriorates for $\Xi<20$, see the right hand side plot where $\Xi=17.5$.}
\label{fig:Xi100final} 
\end{center}
\end{figure}

A new type of the Wigner SC theory of the crystal phase is proposed in 
Sec. \ref{Sec.crystal}, in a perturbative fashion.
At infinite coupling, the counterions stick to the plate surfaces and as 
$d$ increases from 0, they form successively bilayer Wigner crystals 
of rectangular type with the aspect ratio $\Delta$ decreasing
from $\sqrt{3}$ (hexagonal monolayer coined I) to 1 (staggered square 
structure III), see Fig. \ref{fig:Structures}. 
At finite couplings, our MC simulations indicate that counterions are 
still localized around sites of a bilayer structure where 
$\Delta$ depends, besides distance $d$, also on the coupling constant $\Xi$.
In particular, structure I with $\Delta=\sqrt{3}$, which exists only at 
$d=0$ in the ground state, prevails in a nonzero interval of $d$ values 
for finite couplings, see open circles in Fig. \ref{fig:Deltavseta}.
We thus constructed a Wigner-type SC theory based on a harmonic expansion of 
particle coordinates around the sites of the Wigner bilayer, with a free 
aspect ratio $\Delta$, fixed at the end of calculations by minimizing 
the free energy.
Two variants of the WSC expansion were obtained. 
The leading-order one is characterized by an effective one-body potential 
$\kappa \widetilde{z}$ where the prefactor function $\kappa$, which is
$\Delta$-dependent, vanishes for $d\to 0$ (two-plates problem at small distance)
and goes to unity for $\eta\to\infty$ (two separated one-plate problems).
The second variant involves the first correction term $\propto 1/\sqrt{\Xi}$
and, in general, improves substantially the results of the leading-order 
version, even for extremely large coupling constants. 
We have reported a good agreement with Monte Carlo simulation results,
be it for the interplate pressure, or for the ionic density profiles. 
This is the case, expectedly, at very large coupling parameters, where 
the system becomes a (bilayer) Wigner crystal as assumed in our treatment. 
Yet, the predicted pressures and profiles also appear to be reliable at much 
smaller $\Xi$ values, where crystals are completely melted. 
We illustrate this point in Fig. \ref{fig:Xi100final}, where $\Xi=50$ and $\Xi=100$,  
well below the coupling constant of the crystal-fluid transition 
($\Xi$ on the order of 30 000).

Guided by the structure of the WSC results, we also derived a strong-coupling 
description of the fluid regime in Sec. \ref{Sec.fluid}.
Here, the lattice representation of the effective field $\kappa$ 
is replaced by the continuum one based on the idea of a correlation hole,
that has already proven useful in related contexts \cite{Samaj16,Ivan}. 
We proposed two phenomenological constructions of $\kappa$,
with the results (\ref{kappafluid1}) and (\ref{kappafluid2}); 
the corresponding correlation-hole theories were coined as ch1 and ch2.
As is seen in Fig. \ref{fig:kappa}, the approximate plot of $\kappa$
on distance depends only slightly on the choice of the correlation-hole theory,
and ends up close to the WSC derivation.
The correlation hole program leads to observables like pressure and densities
that fare reasonably against numerical simulations. 
Fig. \ref{fig:Xi100final} shows that the ch2 form (slightly better for the chosen 
couplings than ch1), performs as well as the WSC method, while its effective
field $\kappa$ is simpler to compute (compare Eqs. \eqref{kappa} and \eqref{kappafluid2}).
%Yet, even in the fluid regime, which is the one of greatest interest given 
%the very large couplings for crystallization, the WSC approach seems 
%the most efficient, as epitomized in Fig. \ref{fig:Xi100final}. 
%We emphasize that the corresponding analytical formulas are simple as far as 
%the leading order contribution is concerned. 
This completes our goal, since our approach allows to reach distances 
(in Gouy-Chapman units) of order $\sqrt\Xi$, i.e. the typical scale of 
inter-ionic distance. 
To put it differently, no analytical theory could so far account for 
the increasing part of the pressure profile (see e.g. 
Fig. \ref{fig:Xi100final}, or the other pressure plots in this paper)
after the pressure minimum. 
Previous theories \cite{Moreira02,Samaj11} did only describe well 
the decreasing branch of the pressure curve, located at smaller separations.
We recall that the large-distance regime is accounted for by the PB mean-field
theory. 
The latter is repulsive, meaning that the pressure should vanish at 
a specific large but finite distance, a phenomenon that is particularly 
difficult to study analytically, and beyond our scope here. 

In our treatment, we considered as eligible WSC structures rectangle types 
of phases only.
As is seen in Fig. \ref{fig:Deltavseta}, the decrease of the coupling constant 
$\Xi$ increases the $\eta$-range where the structures I-II (and also III)
prevail.
In the ground state, at intermediate to large distances, two different 
structures (staggered rhombic, so-called structure IV, and staggered hexagonal,
structure V) were also observed.
These candidates {\it a priori} impinge on the large-distance WSC calculations 
(i.e. for $\eta$ value of order unity and beyond), but presumably in a modest 
way.
Finally, future plans include extending the present SC methods to 
asymmetrically charged planes and to ions having some structure, 
starting with a hard core. 
A difficulty for the former problem lies in the extreme complexity of 
the ground-state phase diagram \cite{Moritz16}.
Another venue concerns the inclusion of salt (microions with charges of 
both signs).

\begin{acknowledgments}
We would like to dedicate this work to the memory of Per Linse who was 
a true expert in both electrostatics and computer simulations. 
M.T. especially wants to honor Per Linse for having been a great teacher and 
a dear colleague, who always showed a genuine interest in other people's work.
This work was supported by the Grant VEGA  No. 2/0003/18 and by 
the European Union's Horizon 2020 research and innovation programme under 
ETN grant 674979-NANOTRANS.
M.T. acknowledges financial support by the Swedish Research Council
(621-2014-4387).
\end{acknowledgments}

%%%%%%%%%%%%%%%%%%%%%%%%%%%%%%%%%%%%%%%%%%%%%%%%%%%%%%%%%%%%%%%%%%%%%%%

\newpage

\begin{center}
{\bf \Large Supplementary material for ``Strong-coupling theory of counterions 
between symmetrically charged walls: from crystal to fluid phases''}
\end{center}

\vskip 2ex

\setcounter{section}{0}

\renewcommand{\thesection}{\Alph{section}}

We present below some results from previous work for self-containedness
(sections \ref{app:B} and \ref{app:A}),
and calculations explaining the results presented in the main text (sections
\ref{Ssec:harmonic} and \ref{app:C}).
Section \ref{Ssec:harmonic} reports the bulk of our analysis.
We start by ground state features, before working out the harmonic 
expansion treatment leading to the free energy in the crystal phase,
from which thermodynamic properties and ionic profiles follow.

\section{Series representations of the ground-state energy} \label{app:B}
Taking the particle at point $(0,0)$ of plate 1 as a reference, the Coulomb 
interaction energy per particle of structures I-III can be written as 
\begin{equation} \label{e0lattice}
e_0(\eta,\Delta) = \frac{e^2}{2\epsilon a} 
\sum_{(i_x,i_y)\ne (0,0)} \frac{1}{\sqrt{i_x^2+\Delta^2 i_y^2}} + 
\frac{e^2}{2\epsilon a} \sum_{i_x,i_y} 
\frac{1}{\sqrt{\left(i_x-\frac{1}{2}\right)^2 +
\Delta^2 \left(i_y-\frac{1}{2}\right)^2+\left(\frac{d}{a}\right)^2}} 
+ {\rm background} ,
\end{equation}
where the first sum corresponds to the interactions with particles on 
the same plate 1 and the second sum with particles on plate 2.
The background term cancels an infinite constant due to the slow decay
of the Coulomb potential at large distances.

The energy can be reexpressed in terms of a rapidly converging series
by using the method presented in Ref. \cite{Samaj12a}.
We rewrite the ground-state energy per particle as in Eq. (\ref{e0}).
First, using the gamma identity
\begin{equation}
\frac{1}{z^{\nu/2}} = \frac{1}{\Gamma(\nu/2)} \int_0^{\infty} {\rm d}t\,
t^{\frac{\nu}{2}-1} {\rm e}^{-z t}
\end{equation}
($\Gamma$ denotes the Gamma function) with $\nu=1$, 
the $\Sigma$-function is expressed in terms of
Jacobi theta functions with zero argument \cite{Gradshteyn} 
$\theta_3(q) = \sum_j q^{j^2}$ and 
$\theta_2(q) = \sum_j q^{\left(j-\frac{1}{2}\right)^2}$ as follows
\begin{eqnarray}
\Sigma(\eta,\Delta) & = & \int_0^{\infty} \frac{{\rm d}t}{\sqrt{t}} \left\{
\left[ \theta_3\left({\rm e}^{-t\Delta}\right) 
\theta_3\left({\rm e}^{-t/\Delta}\right) - 1 - \frac{\pi}{t} \right] 
\right. \nonumber \\ & & \left.
+ {\rm e}^{-\eta^2 t} \left[ \theta_2\left({\rm e}^{-t\Delta}\right) 
\theta_2\left({\rm e}^{-t/\Delta}\right) - \frac{\pi}{t} \right] \right\} .
\label{Sigma}
\end{eqnarray}
Here, the effect of the background charge density on the plates is
to subtract the singularity $\pi/t$ of the product of theta functions
as $t\to 0$.
Using the Poisson summation formula
\begin{equation}
\sum_{j=-\infty}^{\infty} {\rm e}^{-(j+\phi)^2 t}  = \sqrt{\frac{\pi}{t}}
\sum_{j=-\infty}^{\infty} {\rm e}^{2\pi {\rm i}j\phi} {\rm e}^{-(\pi j)^2/t} ,
\end{equation}
one can reduce the integration support to $t\in [0,\pi]$. 
Applying then once more the Poisson summation formula,
the $\Sigma$-function can be expressed as a series in the generalized
Misra functions (\ref{znu}):
\begin{eqnarray}
\Sigma(\eta,\Delta) & = & 4 \sum_{j=1}^{\infty} \left[
z_{3/2}\left(0,j^2/\Delta\right) + z_{3/2}\left(0,j^2\Delta\right) \right]
+ 8 \sum_{j,k=1}^{\infty} z_{3/2}\left(0,j^2/\Delta+k^2\Delta\right) \nonumber \\
& & + 2 \sum_{j=1}^{\infty} (-1)^j \left[
z_{3/2}\left((\pi\eta)^2,j^2/\Delta\right) + 
z_{3/2}\left((\pi\eta)^2,j^2\Delta\right) \right] 
+ 4 \sum_{j,k=1}^{\infty} (-1)^j (-1)^k 
z_{3/2}\left((\pi\eta)^2,j^2/\Delta+k^2\Delta\right) \nonumber \\
& & + 4 \sum_{j,k=1}^{\infty} 
z_{3/2}\left(0,\eta^2+(j-1/2)^2/\Delta+(k-1/2)^2\Delta\right) - 4\sqrt{\pi}
-\pi z_{1/2}(0,\eta^2) . \label{sigmaMisra} 
\end{eqnarray}

\section{Generalized Misra functions} \label{app:A}
The first few generalized Misra functions $z_{\nu}(x,y)$ (\ref{znu}) with 
half-integer arguments are expressible in terms of the complementary error 
function \cite{Gradshteyn}
\begin{equation} \label{erfc}
{\rm erfc}(u)=\frac{2}{\sqrt{\pi}}\int_u^\infty\exp{(-t^2)}\ {\rm d}t ,
\end{equation}
as follows \cite{Travenec15}:
\begin{eqnarray} 
z_{1/2}(x,y) & = & \sqrt{\frac{\pi}{x}}{\rm e}^{-2\sqrt{xy}}
\left[ 1-\frac{1}{2}\ {\rm erfc}{\left(\sqrt{\frac{x}{\pi}}-\sqrt{\pi y}\right)}
-\frac{1}{2}{\rm e}^{4\sqrt{xy}}\ 
{\rm erfc}{\left(\sqrt{\frac{x}{\pi}}+\sqrt{\pi y}\right)}\right], 
\nonumber \\
z_{3/2}(x,y) & = & \sqrt{\frac{\pi}{y}}{\rm e}^{-2\sqrt{xy}}\left[ 1-\frac{1}{2}\ 
{\rm erfc}{\left(\sqrt{\frac{x}{\pi}}-\sqrt{\pi y}\right)}
+ \frac{1}{2}{\rm e}^{4\sqrt{xy}}\ 
{\rm erfc}{\left(\sqrt{\frac{x}{\pi}}+\sqrt{\pi y}\right)}\right], 
\nonumber \\
z_{5/2}(x,y) & = & \frac{\sqrt{\pi x}}{y}{\rm e}^{-2\sqrt{xy}}
\left(1+\frac{1}{2\sqrt{xy}}\right)
-\frac{\sqrt{\pi}}{4y^{3/2}}\bigg[-4{\rm e}^{-x/\pi-\pi y}\sqrt{y} 
\nonumber \\ & & 
+{\rm e}^{-2\sqrt{xy}}\left(1+2\sqrt{xy}\right)\ 
{\rm erfc}{\left(\sqrt{\frac{x}{\pi}}-\sqrt{\pi y}\right)}
+{\rm e}^{2\sqrt{xy}}\left(-1+2\sqrt{xy}\right)\ 
{\rm erfc}{\left(\sqrt{\frac{x}{\pi}}+\sqrt{\pi y}\right)} \bigg] . 
\label{zerror}
\end{eqnarray}
The case of the ordinary Misra functions $z_{\nu}(0,y)$ \cite{Misra} should be 
understood in the sense of the limit $x\to 0$,
\begin{eqnarray}
z_{1/2}(0,y) & = &\frac{2}{\sqrt{\pi}}\left[{\rm e}^{-\pi y} -\pi \sqrt{y} \ 
{\rm erfc}{\left(\sqrt{\pi y}\right)}\right], \nonumber\\
z_{3/2}(0,y) & = &\sqrt{\frac{\pi}{y}}\ {\rm erfc}{\left(\sqrt{\pi y}\right)},
\nonumber\\
z_{5/2}(0,y) & = &\frac{\sqrt{\pi}}{2 y^{3/2}}\left[2 {\rm e}^{-\pi y}\sqrt{y} \ 
+{\rm erfc}{\left(\sqrt{\pi y}\right)}\right] . \label{znu0y}
\end{eqnarray}

\section{Large-coupling description of the crystal phase} \label{Ssec:harmonic}

\subsection{Harmonic expansion of the energy} \label{Sssec:harmonic}
Starting from a crystalline configuration, let us shift each particle $i$ 
at plate $\Sigma_1$ from its reference Wigner-lattice position 
(\ref{partplate1}) to 
\begin{equation} \label{shiftpartplate1}
{\bf r}_i = (a i_x + x_i,\Delta a i_y + y_i,z_i) ,
\end{equation}
where the coordinate shifts $x_i$, $y_i$ and $z_i$ are assumed to be small.
Similarly, we shift the Wigner position (\ref{partplate2}) of each particle 
at plate $\Sigma_2$ to the one
\begin{equation} \label{shiftpartplate2}
{\bf r}_i = \left( a \left( i_x-\frac{1}{2}\right)+x_i,
\Delta a \left( i_y-\frac{1}{2}\right)+y_i,d-(d-z_i)\right) ,
\end{equation}
where now $x_i$, $y_i$ and $d-z_i$ are assumed to be small.

If the particles $i\to (i_x,i_y)$ and $j\to (j_x,j_y)$ are localized 
on the same plate, either $\Sigma_1$ or $\Sigma_2$, the change of 
the Coulomb energy due to the particle shifts reads as
\begin{equation}
\delta E_{ij} = \frac{e^2}{\epsilon} \left[
\frac{1}{\sqrt{[a(i_x-j_x)+(x_i-x_j)]^2+[a\Delta (i_y-j_y)+(y_i-y_j)]^2+
(z_i-z_j)^2}} - \frac{1}{\sqrt{a^2(i_x-j_x)^2+a^2\Delta^2 (i_y-j_y)^2}} \right] .
\end{equation}
If both particles are at plate $\Sigma_1$, the expansion of $\delta E_{ij}$
in small deviations $(x_i,x_j)$, $(y_i,y_j)$ and $(z_i,z_j)$ is straightforward.
Since $z_i-z_j \equiv (d-z_j) - (d-z_i)$, the same holds for two particles
being at plate $\Sigma_2$ where the deviations $d-z_i$ and $d-z_j$ are small.
If particles $i$ and $j$ belong to different plates, say $i\in \Sigma_1$ 
and $j\in\Sigma_2$, the energy change is given by 
\begin{eqnarray}
\delta E_{ij} & = & \frac{e^2}{\epsilon} \left[
\frac{1}{\sqrt{[a(i_x-j_x-1/2)+(x_i-x_j)]^2+[a\Delta (i_y-j_y-1/2)+(y_i-y_j)]^2+
(z_i-z_j)^2}} \right. \nonumber \\ & & \left.
- \frac{1}{\sqrt{a^2(i_x-j_x-1/2)^2+a^2\Delta^2 (i_y-j_y-1/2)^2+d^2}} \right] .
\end{eqnarray}
In this case, we write $z_i-z_j \equiv -d + z_i + (d-z_j)$ and perform
the expansion of the energy change in the small quantities $z_i$ and $(d-z_j)$.
The total energy is expressible as
\begin{equation} 
E(\{ {\bf r}_i\}) = N e_0(\eta,\Delta) + \delta E , \qquad
\delta E = \sum_{i<j} \delta E_{ij} .
\end{equation}
Within the harmonic approximation, we expand every $\delta E_{ij}$ up to 
quadratic terms in small deviations, supposing that the ratios 
$x_i/a$, $y_i/a$, $z_i/a$ are small variables for particles $i\in \Sigma_1$ 
and that $x_i/a$, $y_i/a$, $(d-z_i)/a$ are small variables for particles 
$i\in \Sigma_2$. 
Many terms disappear because of the symmetry of the energy with respect to 
the reflection transformations $x\to -x$ and $y\to -y$.
The final result for the energy change is Eq. (\ref{deltaE}) in the main text:
\begin{equation}
- \beta \delta E = - \kappa(\eta,\Delta) 
\left[ \sum_{i\in\Sigma_1} \widetilde{z}_i + \sum_{i\in\Sigma_2} 
(\widetilde{d}-\widetilde{z}_i) \right] + \frac{1}{\sqrt{\Xi}} S_z
- \sqrt{\frac{\Xi}{2\pi}} \frac{\sigma}{2} \sum_{i<j}
\left[ B_{ij}^x (x_i-x_j)^2 + B_{ij}^y (y_i-y_j)^2 \right] + \cdots ,
\end{equation}
with
\begin{eqnarray}
\kappa(\eta,\Delta) & = & \frac{\eta}{2\pi} \sum_{i_x,i_y} 
\frac{\Delta^{3/2}}{\left[ (i_x-1/2)^2 + \Delta^2 (i_y-1/2)^2
+\Delta\eta^2\right]^{3/2}} \nonumber \\
& = & \frac{\eta}{\pi^{3/2}} \int_0^{\infty} {\rm d}t\, \sqrt{t}\, 
{\rm e}^{-\eta^2 t} \theta_2\left({\rm e}^{-t\Delta}\right) 
\theta_2\left({\rm e}^{-t/\Delta}\right) = 
- \frac{1}{2 \pi^{3/2}} \frac{\partial}{\partial\eta} \Sigma(\eta,\Delta) + 1 .
\end{eqnarray}
The quantity $S_z$ involves all terms quadratic in variables $\widetilde{z}_i$
if $i\in \Sigma_1$ and $(\widetilde{d}-\widetilde{z}_i)$ if $i\in \Sigma_2$, 
\begin{eqnarray}
S_z(\eta,\Delta) & = & \frac{\Delta^{3/2}}{2(2\pi)^{3/2}}
\Bigg\{ \sum_{i,j\in\Sigma_1 \atop (i<j)} 
\frac{1}{\left[ (i_x-j_x)^2 + \Delta^2(i_y-j_y)^2\right]^{3/2}}
\left( \widetilde{z}_i^2 + \widetilde{z}_j^2 - 2 \widetilde{z}_i 
\widetilde{z}_j \right) \nonumber \\ & & + \sum_{i,j\in\Sigma_2 \atop (i<j)} 
\frac{1}{\left[ (i_x-j_x)^2 + \Delta^2(i_y-j_y)^2\right]^{3/2}}
\left[ (\widetilde{d}-\widetilde{z}_i)^2 + (\widetilde{d}-\widetilde{z}_j)^2 - 
2 (\widetilde{d}-\widetilde{z}_i)(\widetilde{d}-\widetilde{z}_j) \right] 
\nonumber \\ & &
+ \sum_{i\in\Sigma_1} \sum_{j\in\Sigma_2} 
\left[ \frac{1}{\left[ (i_x-j_x-1/2)^2 + \Delta^2(i_y-j_y-1/2)^2+
\Delta \eta^2\right]^{3/2}} \right. \nonumber \\ & & \left.
- \frac{3\Delta\eta^2}{\left[ 
(i_x-j_x-1/2)^2 + \Delta^2(i_y-j_y-1/2)^2+\Delta \eta^2\right]^{5/2}} \right]
\left[ \widetilde{z}_i^2 + (\widetilde{d}-\widetilde{z}_j)^2 +
2 \widetilde{z}_i(\widetilde{d}-\widetilde{z}_j) \right] \Bigg\}  \label{Sz}
\end{eqnarray}
and the expansion coefficients in the $(x,y)$-plane are given by
\begin{eqnarray} 
B_{ij}^x(\Delta) & = & \Delta^{3/2} \frac{2 (i_x-j_x)^2-\Delta^2(i_y-j_y)^2}{
\left[ (i_x-j_x)^2 + \Delta^2(i_y-j_y)^2\right]^{5/2}} , \nonumber \\
B_{ij}^y(\Delta) & = & \Delta^{3/2} \frac{2 \Delta^2 (i_y-j_y)^2 - (i_x-j_x)^2}{
\left[ (i_x-j_x)^2 + \Delta^2(i_y-j_y)^2\right]^{5/2}} \label{Bsame}
\end{eqnarray}
if particles $i$ and $j$ belong to the same plate and by
\begin{eqnarray} 
B_{ij}^x(\eta,\Delta) & = & \Delta^{3/2} \frac{2 (i_x-j_x-1/2)^2
-\Delta^2(i_y-j_y-1/2)^2-\Delta\eta^2}{\left[ (i_x-j_x-1/2)^2 
+ \Delta^2(i_y-j_y-1/2)^2+\Delta\eta^2 \right]^{5/2}} , \nonumber \\
B_{ij}^y(\eta,\Delta) & = & \Delta^{3/2} \frac{2 \Delta^2 (i_y-j_y-1/2)^2 
- (i_x-j_x-1/2)^2-\Delta\eta^2}{\left[ (i_x-j_x-1/2)^2 
+ \Delta^2(i_y-j_y-1/2)^2+\Delta\eta^2 \right]^{5/2}} \label{Bdifferent} 
\end{eqnarray}
if particles $i$ and $j$ belong to different plates.

\subsection{Thermodynamics} \label{Sssec:thermo}
To express $\ln Q_z$ as a perturbative series in $S_z$, we introduce 
the counterpart of (\ref{Qz}) for non-interacting $(S_z=0)$ 
particles in the external potential only:
\begin{equation} \label{Qzp}
Q_z^{(0)}(\eta,\Delta) = \int_0^{\widetilde{d}} \prod_{i\in\Sigma_1} 
{\rm d}\widetilde{z}_i\, {\rm e}^{-\kappa \widetilde{z}_i} 
\int_0^{\widetilde{d}} \prod_{i\in\Sigma_2} {\rm d}\widetilde{z}_i\,
{\rm e}^{-\kappa(\widetilde{d}-\widetilde{z}_i)} 
= \left( \frac{1 - \exp(-\kappa\widetilde{d})}{\kappa} \right)^N .
\end{equation}
We have
\begin{equation}
\ln \left( \frac{Q_z}{Q_z^{(0)}} \right) = \ln \langle \exp(S_z)\rangle_0 ,
\end{equation}
where $\langle \cdots \rangle_0$ denotes the statistical averaging over 
the system of non-interacting particles defined by the partition sum 
$Q_z^{(0)}$. 
The quantity $\ln \langle \exp(S_z)\rangle_0$ can be written as
the cumulant expansion:
\begin{equation}
\ln \langle \exp(S_z)\rangle_0 = \langle S_z \rangle_0 +
\frac{1}{2!} \left( \langle S_z^2 \rangle_0 - \langle S_z \rangle_0^2 \right)
+ \cdots ,
\end{equation}
where each term of the expansion is extensive, i.e. proportional to 
the particle number $N$.
Restricting ourselves to the lowest cumulant order, we obtain
\begin{equation}
\frac{1}{N} \ln Q_z = \ln \left( \frac{1 - \exp(-\kappa\widetilde{d})}{\kappa} 
\right) + \frac{1}{N} \langle S_z \rangle_0 
\end{equation}
with 
$\widetilde{d} \equiv d/\mu = \eta/(\mu\sqrt{\sigma}) = \eta\sqrt{2\pi\Xi}$. 
The evaluation of $\langle S_z \rangle_0/N$ yields:
\begin{eqnarray} 
\frac{1}{N} \langle S_z \rangle_0 & = & \frac{1}{2(2\pi)^{3/2}}
\left\{ \frac{F(\Delta)}{2} \left[ 
\left( \langle \widetilde{z}^2 \rangle_0 - \langle \widetilde{z} \rangle_0^2 
\right) + \left( \langle (\widetilde{d}-\widetilde{z})^2 \rangle_0 - 
\langle (\widetilde{d}-\widetilde{z}) \rangle_0^2 \right) \right] \right.
\nonumber \\ & & \left.
+ \pi \frac{\partial\kappa(\eta,\Delta)}{\partial\eta} 
\left[ \langle \widetilde{z}^2 \rangle_0 +\langle 
(\widetilde{d}-\widetilde{z})^2 \rangle_0  + 2 \langle \widetilde{z} \rangle_0 
\langle (\widetilde{d}-\widetilde{z}) \rangle_0 \right] \right\} , \label{Sz0}
\end{eqnarray}
where $F(\Delta)$ corresponds to the lattice sum
\begin{equation} \label{F} 
F(\Delta) = \sum_{(i_x,i_y)\ne (0,0)} 
\frac{\Delta^{3/2}}{(i_x^2+\Delta^2i_y^2)^{3/2}}
\end{equation} 
and the one-body averages 
\begin{equation}
\langle \widetilde{z}^p \rangle_0 = 
\frac{\int_0^{\widetilde{d}} {\rm d}\widetilde{z}\,
\widetilde{z}^p {\rm e}^{-\kappa\widetilde{z}}}{\int_0^{\widetilde{d}} 
{\rm d}\widetilde{z}\, {\rm e}^{-\kappa\widetilde{z}}} , \qquad
\langle (\widetilde{d}-\widetilde{z})^p \rangle_0 = 
\frac{\int_0^{\widetilde{d}} {\rm d}\widetilde{z}\,
(\widetilde{d}-\widetilde{z})^p {\rm e}^{-\kappa(\widetilde{d}-\widetilde{z})}}{
\int_0^{\widetilde{d}} {\rm d}\widetilde{z}\, 
{\rm e}^{-\kappa(\widetilde{d}-\widetilde{z})}} .
\end{equation}
In particular, we shall need
\begin{eqnarray}
\langle \widetilde{z} \rangle_0 & = & \langle (\widetilde{d}-\widetilde{z}) 
\rangle_0  = \frac{1}{\kappa} - 
\frac{\widetilde{d}}{{\rm e}^{\kappa\widetilde{d}}-1} , \\ 
\langle \widetilde{z}^2 \rangle_0 & = & \langle 
(\widetilde{d}-\widetilde{z})^2 \rangle_0  
= \frac{2}{\kappa^2} - \frac{\widetilde{d}(2+\kappa\widetilde{d})}{\kappa
\left({\rm e}^{\kappa\widetilde{d}}-1\right)} .
\end{eqnarray}

To calculate the integral $Q_x$ in (\ref{Qx}), we respect the $x$-coordinate
constraint (\ref{cell}) and rescale the particle $x$-coordinates by 
the factor $(2\pi/\Xi)^{1/4}/\sqrt{\sigma}$ to obtain
\begin{equation} \label{Qx1}
Q_x(\eta,\Delta) = \left( \frac{2\pi}{\Xi} \right)^{N/4}
\frac{1}{(\lambda\sqrt{\sigma})^N} \int_{-L}^L \prod_{i\in\Sigma_1\cup \Sigma_2} 
{\rm d}x_i\, \exp\left[ - \frac{1}{2} \sum_{i<j} B_{ij}^x (x_i-x_j)^2 \right] ,
\end{equation}
where $L\propto \Xi^{1/4}$ goes to infinity in the large-$\Xi$ limit. 
Here, going back to dimensioned lengths, a new relevant length scale arises, 
$a/\Xi^{1/4}$. 
It is readily checked that it measures the amplitude of in plane 
$x$-fluctuations around a lattice position. 
Incidentally, we note first that a similar scaling arises for the minimum of 
the pressure curves, in the regime of like-charge attraction, that is largely 
met here \cite{Chen06,Samaj11}. 
Second, this provides a new light on the melting criterion alluded to above, 
where the critical coupling in the 2D-confined problem is around 15000. 
This yields $a/\Xi^{1/4}\simeq 0.09 a$, a value close to Lindeman type of 
criteria \cite{HansenMcDonald}.
To avoid the divergence of the consequent integral manifesting itself by 
the invariance of $\sum_{i<j} B_{ij}^x (x_i-x_j)^2 $ with respect to a uniform 
coordinate shift $x_i\to x_i+c$, we shall make provision for finiteness of 
the $L$-bound for a large but finite $\Xi$ and ignore the zero Fourier mode, 
see below.
Omitting in (\ref{Qx1}) irrelevant prefactors we end up with the integral of 
Gaussian type
\begin{equation}
Q_x(\eta,\Delta) = \int_{-\infty}^{\infty} \prod_{i=1}^N 
{\rm d}x_i\, \exp\left[ - \frac{1}{2} \sum_{i,j=1}^N A_{ij}^x x_i x_j \right] 
=  \frac{(2\pi)^{N/2}}{\sqrt{\text{Det}\,{\bf A}^x}} ,
\end{equation}
where the ${\bf A}^x$-matrix is defined by
\begin{equation} \label{Ax}
A_{ii}^x = \sum_{k\ne i} B_{ik}^x , \qquad A_{ij}^x = - B_{ij}^x \quad
\mbox{for $(i\ne j)$.}
\end{equation}

According to Fig. \ref{fig:Structures}, within the $(x,y)$ plane we can 
represent the Wigner bilayer as the regular 2D lattice of alternating white 
(belonging to plate $\Sigma_1$) and black (belonging to $\Sigma_2$) points, 
with the primitive translation vectors
\begin{equation}
\bm{\alpha} = a (1,0) , \qquad {\bm\beta} = \frac{a}{2} (1,\Delta) 
\end{equation}    
and the surface of the elementary cell $S=\Delta a^2/2$.
The matrix elements $A_{ij}^x$ depend only on the distance of lattice
points $i,j$ and therefore ${\bf A}^x$ is an $N\times N$ circulant matrix 
with known eigenvalue spectrum.
Let us define the 2D Fourier transform of any lattice function 
$h_{ij}=f(\vert {\bf r}_i-{\bf r}_j\vert)$ as follows 
\begin{equation}
h({\bf q}) = \sum_k h_{jk} {\rm e}^{{\rm i}{\bf q}\cdot ({\bf r}_j-{\bf r}_k)} ,
\end{equation}
where the $N$ vectors ${\bf q}=(q_x,q_y)$ belong to the first Brillouin zone 
(BZ) of the reciprocal lattice with the primitive vectors ${\bm\alpha}^*$,
${\bm\beta}^*$ defined by the relations
\begin{equation}
{\bm\alpha}^*\cdot {\bm\alpha} = {\bm\beta}^*\cdot {\bm\beta} = 2\pi, \qquad
{\bm\alpha}^*\cdot {\bm\beta} = {\bm\alpha}\cdot {\bm\beta}^* = 0 .
\end{equation}
In particular,
\begin{equation} \label{vectorsBZ}
{\bm\alpha}^* = \frac{2\pi}{a} \left( 1,-\frac{1}{\Delta} \right) , \qquad 
{\bm\beta}^* = \frac{4\pi}{a} \left( 0,\frac{1}{\Delta} \right)
\end{equation}
and the surface of the BZ is given by $S^* = 8\pi^2/(\Delta a^2)$. 
Since the $A^x({\bf q})$ with ${\bf q}\in\text{BZ}$ are the $N$ eigenvalues  
of the matrix ${\bf A}^x$, we have 
\begin{equation}
\text{Det}\,{\bf A}^x = \prod_{{\bf q}\in\text{BZ}\atop {\bf q}\ne 0} A^x({\bf q}) ,
\qquad -\frac{1}{N} \ln Q_x(\eta,\Delta) = \frac{1}{2N} 
\sum_{{\bf q}\in\text{BZ}\atop {\bf q}\ne 0} \ln A^x({\bf q}),
\end{equation}
the zero-mode being excluded.
In the thermodynamic limit $N\to\infty$, the ${\bf q}$-vectors cover
uniformly the BZ defined by the primitive vectors (\ref{vectorsBZ}) and 
we can write
\begin{eqnarray}
\frac{1}{N} \sum_{{\bf q}\in\text{BZ}\atop {\bf q}\ne 0} \ln A^x({\bf q}) & = & 
\frac{1}{S^*} \int_{\text{BZ}} {\rm d}{\bf q}\, \ln A^x({\bf q}) = 
\frac{\Delta a^2}{8\pi^2}
\int_0^{2\pi/a}{\rm d}q_x \int_{-q_x/\Delta}^{4\pi/(a\Delta)-q_x/\Delta} {\rm d}q_y\,
\ln A^x(q_x,q_y) \nonumber \\ & = &
\frac{1}{2} \int_0^1 {\rm d}q_x \int_0^2 {\rm d}q_y\,
\ln A^x\left[ \frac{2\pi}{a} q_x,\frac{2\pi}{a\Delta} (q_y-q_x) \right] .
\end{eqnarray}
Consequently,
\begin{equation}
- \lim_{N\to\infty} \frac{1}{N} \ln Q_x(\eta,\Delta) = 
\frac{1}{4} \int_0^1 {\rm d}q_x \int_0^2 {\rm d}q_y\,
\ln A^x\left[ \frac{2\pi}{a} q_x,\frac{2\pi}{a\Delta} (q_y-q_x) \right] .
\end{equation}

Now we want to express appropriately the Fourier component  
$A^x(2\pi q_x/a,2\pi q_y/(a\Delta)$, the elements of the ${\bf A}^x$-matrix 
being defined in terms of those of the ${\bf B}^x$-matrix   
[see formulas (\ref{Bsame}) and (\ref{Bdifferent})] in Eq. (\ref{Ax}).
We introduce the auxiliary Fourier lattice functions
\begin{eqnarray}
F(\Delta;{\bf q}) & = & \sum_{(i_x,i_y)\ne (0,0)} 
\frac{\Delta^{3/2}}{(i_x^2+\Delta^2i_y^2)^{3/2}}
{\rm e}^{{\rm i}2\pi(q_x i_x + q_y i_y)} , \label{Fq} \\
G(\eta,\Delta;{\bf q}) & = & \sum_{(i_x,i_y)} 
\frac{\Delta^{3/2}}{((i_x-1/2)^2+\Delta^2(i_y-1/2)^2+\Delta\eta^2)^{3/2}}
{\rm e}^{{\rm i}2\pi[q_x (i_x-1/2) + q_y (i_y-1/2)]}  . \label{Gq}
\end{eqnarray}
Note that the previous lattice sum (\ref{F}) is expressible as 
$F(\Delta)=F(\Delta,{\bf 0})$. 
The Misra series representations of $F(\Delta;{\bf q})$ and 
$G(\eta,\Delta;{\bf q})$ are given in Eqs. (\ref{Fseries}) and
(\ref{Gseries}) in section \ref{app:C}, respectively.
Introducing the function
\begin{eqnarray}
C^x({\bf q}) & = & \frac{1}{2} F(\Delta;{\bf q}) + 
\Delta \frac{\partial}{\partial\Delta} F(\Delta;{\bf q}) +
\frac{1}{2} G(\eta,\Delta;{\bf q}) \nonumber \\ & & + 
\Delta \frac{\partial}{\partial\Delta} G(\eta,\Delta;{\bf q}) +
\frac{\eta}{2} \frac{\partial}{\partial\eta} G(\eta,\Delta;{\bf q})
\label{eq:cx}
\end{eqnarray}
it holds that
\begin{equation}
A^x\left( \frac{2\pi}{a} q_x,\frac{2\pi}{a\Delta} q_y \right)
= C^x(0,0) - C^x(q_x,q_y) .
\end{equation}

To evaluate the integral $Q_y$ (\ref{Qy}), we proceed analogously.
The ${\bf A}^y$-matrix is defined by
\begin{equation} \label{Ay}
A_{ii}^y = \sum_{k\ne i} B_{ik}^y , \qquad A_{ij}^y = - B_{ij}^y \quad
\mbox{for $(i\ne j)$,}
\end{equation}
see Eqs. (\ref{Bsame}) and (\ref{Bdifferent}) for the ${\bf B}^y$-matrix
elements.
In the thermodynamic limit we find that    
\begin{equation}
- \lim_{N\to\infty} \frac{1}{N} \ln Q_y(\eta,\Delta) = 
\frac{1}{4} \int_0^1 {\rm d}q_x \int_0^2 {\rm d}q_y\,
\ln A^y\left[ \frac{2\pi}{a} q_x,\frac{2\pi}{a\Delta} (q_y-q_x) \right] .
\end{equation}
Here,
\begin{equation}
A^y\left( \frac{2\pi}{a} q_x,\frac{2\pi}{a\Delta} q_y \right)
= C^y(0,0) - C^y(q_x,q_y) ,
\end{equation}
where the auxiliary function
\begin{eqnarray}
C^y({\bf q}) & = & \frac{1}{2} F(\Delta;{\bf q}) - 
\Delta \frac{\partial}{\partial\Delta} F(\Delta;{\bf q}) +
\frac{1}{2} G(\eta,\Delta;{\bf q}) \nonumber \\ & & - 
\Delta \frac{\partial}{\partial\Delta} G(\eta,\Delta;{\bf q}) +
\frac{\eta}{2} \frac{\partial}{\partial\eta} G(\eta,\Delta;{\bf q}) .
\label{eq:cy}
\end{eqnarray}

%%%%%%%%%%%%%%%%%%%%%%%%%%%%%%%%%%%%%%%%%%%%%%%%%%%
\subsection{Particle density profile and pressure}
\label{Sssec:density}

We start from
\begin{equation} 
Z_N[w] = \frac{1}{N!} \int_{\Lambda} \prod_{i=1}^N \frac{{\rm d}{\bf r}_i}{
\lambda^3}\, w({\bf r}_i) {\rm e}^{-\beta E(\{ {\bf r}_i\})} ,
\end{equation} 
a functional of the generating Boltzmann weight
$w({\bf r}) = \exp[-\beta u({\bf r})]$, such that
\begin{equation}
\rho({\bf r}) = \frac{\delta}{\delta w({\bf r})} \ln Z_N[w] 
\Big\vert_{w({\bf r})=1} .
\end{equation}
For our $z$-dependent density $\rho(z)$ one can ignore harmonic modes along 
the $(x,y)$ plane as well as $w$-independent terms.
After simple algebra, we find that
\begin{equation}
\ln Z_N[w] = \frac{N}{2} \ln \left[ \int_{\Lambda} {\rm d}{\bf r}\, w({\bf r})
{\rm e}^{-\kappa\widetilde{z}} \right] + 
\frac{N}{2} \ln \left[ \int_{\Lambda} {\rm d}{\bf r}\, w({\bf r})
{\rm e}^{-\kappa(\widetilde{d}-\widetilde{z})} \right] 
+ \frac{1}{\sqrt{\Xi}} \langle S_z[w] \rangle_0 ,
\end{equation}
where the functional $\langle S_z[w] \rangle_0$ is given by Eq.
(\ref{Sz0}) with the moments redefined as follows
\begin{eqnarray}
\langle \widetilde{z}^p \rangle_0 & \to & \langle \widetilde{z}^p[w] \rangle_0 
= \frac{\int_{\Lambda} {\rm d}{\bf r}\, 
w({\bf r}) \widetilde{z}^p {\rm e}^{-\kappa\widetilde{z}}}{\int_{\Lambda}
{\rm d}{\bf r}\, w({\bf r}) {\rm e}^{-\kappa\widetilde{z}}} , \nonumber \\
\langle (\widetilde{d}-\widetilde{z})^p \rangle_0 & \to &
\langle (\widetilde{d}-\widetilde{z})^p[w] \rangle_0 = 
\frac{\int_{\Lambda} {\rm d}{\bf r}\, w({\bf r})
(\widetilde{d}-\widetilde{z})^p {\rm e}^{-\kappa(\widetilde{d}-\widetilde{z})}}{
\int_{\Lambda} {\rm d}{\bf r}\, w({\bf r}) {\rm e}^{-\kappa(\widetilde{d}-\widetilde{z})}} .
\end{eqnarray}
Then the (rescaled) particle density can be represented as the WSC expansion
\begin{equation} 
\widetilde{\rho}(\widetilde{z}) = \widetilde{\rho}^{(0)}(\widetilde{z}) 
+ \frac{1}{\sqrt{\Xi}} \widetilde{\rho}^{(1)}(\widetilde{z}) + \cdots .
\end{equation}
Since
\begin{equation}
\frac{\delta}{\delta w({\bf r})} \frac{N}{2} 
\ln \left[ \int_{\Lambda} {\rm d}{\bf r}\, w({\bf r}) 
{\rm e}^{-\kappa\widetilde{z}} \right] \Bigg\vert_{w({\bf r})=1} 
= \frac{N {\rm e}^{-\kappa\widetilde{z}}}{
2\int_{\Lambda}{\rm d}{\bf r}\, {\rm e}^{-\kappa\widetilde{z}}}
= \frac{N\kappa}{2 S \mu\left( 1-{\rm e}^{-\kappa\widetilde{d}}\right)} 
{\rm e}^{-\kappa\widetilde{z}}
\end{equation}
and $N/(2S\mu) = 2\pi\ell_{\rm B}\sigma^2$, we have in the leading WSC order
\begin{equation} 
\widetilde{\rho}^{(0)}(\widetilde{z}) = \frac{\kappa}{1-{\rm e}^{-\kappa\widetilde{d}}}
\left[ {\rm e}^{-\kappa\widetilde{z}} + {\rm e}^{-\kappa(\widetilde{d}-\widetilde{z})} \right] .
\end{equation} 

The first correction to the particle density 
$\widetilde{\rho}^{(1)}(\widetilde{z})$ is generated from $\langle S_z[w] \rangle_0$ 
by using the functional derivatives of the moments
\begin{eqnarray}
\frac{\delta}{\delta w({\bf r})} \langle\widetilde{z}^p[w]\rangle_0 
\Big\vert_{w({\bf r})=1} & = & 
\frac{\kappa}{S\mu\left(1-{\rm e}^{-\kappa\widetilde{d}}\right)}
{\rm e}^{-\kappa\widetilde{z}} \left( \widetilde{z}^p - \langle \widetilde{z}^p\rangle_0 
\right) , \\
\frac{\delta}{\delta w({\bf r})} \langle(\widetilde{d}-\widetilde{z})^p[w]\rangle_0 
\Big\vert_{w({\bf r})=1} & = & 
\frac{\kappa}{S\mu\left(1-{\rm e}^{-\kappa\widetilde{d}}\right)}
{\rm e}^{-\kappa(\widetilde{d}-\widetilde{z})} \left[
(\widetilde{d}-\widetilde{z})^p - \langle (\widetilde{d}-\widetilde{z})^p \rangle_0 \right] .
\label{eq:rho1}
\end{eqnarray}
In particular,
\begin{eqnarray}
\widetilde{\rho}^{(1)}(\widetilde{z}) & = & 
\frac{\kappa}{(2\pi)^{3/2}\left( 1-{\rm e}^{-\kappa\widetilde{d}}\right)} \Bigg\{
F(\Delta) {\rm e}^{-\kappa\widetilde{z}} \left[ 
\frac{\widetilde{z}^2-\langle\widetilde{z}^2\rangle_0}{2} - \langle\widetilde{z}\rangle_0 
\left( \widetilde{z}-\langle\widetilde{z}\rangle_0 \right) \right] \nonumber \\
& & + F(\Delta) {\rm e}^{-\kappa(\widetilde{d}-\widetilde{z})} \left[ 
\frac{(\widetilde{d}-\widetilde{z})^2-\langle(\widetilde{d}-\widetilde{z})^2\rangle_0}{2} 
- \langle(\widetilde{d}-\widetilde{z})\rangle_0 \left( (\widetilde{d}-\widetilde{z})
-\langle(\widetilde{d}-\widetilde{z})\rangle_0 \right) \right] \nonumber \\ & &
+ 2\pi \frac{\partial\kappa(\eta,\Delta)}{\partial\eta}
\Bigg[ {\rm e}^{-\kappa\widetilde{z}} \frac{\widetilde{z}^2-\langle\widetilde{z}^2\rangle_0}{2}
+ {\rm e}^{-\kappa(\widetilde{d}-\widetilde{z})} \frac{(\widetilde{d}-\widetilde{z})^2-
\langle(\widetilde{d}-\widetilde{z})^2\rangle_0}{2} \nonumber \\ & &
+ {\rm e}^{-\kappa\widetilde{z}} \langle(\widetilde{d}-\widetilde{z})\rangle_0
\left( \widetilde{z} - \langle\widetilde{z}\rangle_0 \right)
+ {\rm e}^{-\kappa(\widetilde{d}-\widetilde{z})} \langle\widetilde{z}\rangle_0
\left( (\widetilde{d}-\widetilde{z}) - \langle(\widetilde{d}-\widetilde{z})\rangle_0 \right)
\Bigg] \Bigg\} . \label{rho1}
\end{eqnarray}
Because of the equalities
\begin{equation}
\int_{\Lambda} {\rm d}{\bf r} \frac{\delta}{\delta w({\bf r})} 
\langle\widetilde{z}^p[w]\rangle_0 \Big\vert_{w({\bf r})=1} =
\int_{\Lambda} {\rm d}{\bf r} \frac{\delta}{\delta w({\bf r})} 
\langle(\widetilde{d}-\widetilde{z})^p[w]\rangle_0 \Big\vert_{w({\bf r})=1} = 0 ,
\end{equation}
we have 
\begin{equation} 
\int_0^{\widetilde{d}} {\rm d}\widetilde{z}\, \widetilde{\rho}^{(1)}(\widetilde{z}) = 0 ,
\end{equation}
so that 
the electroneutrality condition is met.

Finally, the contact theorem for planar walls \cite{contact} relates the total
contact density of particles on the wall and the pressure.
Within our notation, it is expressible as
\begin{equation}
\widetilde{P}_{\rm c} = \widetilde{\rho}(0) - 1 = \left[ \widetilde{\rho}^{(0)}(0)-1 \right]
+ \frac{1}{\sqrt{\Xi}} \widetilde{\rho}^{(1)}(0) + \cdots .
\end{equation}
Writing the WSC expansion for the ``contact'' pressure as
$\widetilde{P}_{\rm c} = \widetilde{P}_{\rm c}^{(0)} + 
 \widetilde{P}_{\rm c}^{(1)} / \sqrt{\Xi} + \cdots 
$,
we get
\begin{equation}
\widetilde{P}_c^{(0)} = \kappa \left( \frac{1+{\rm e}^{-\kappa\widetilde{d}}}{
1-{\rm e}^{-\kappa\widetilde{d}}} \right) - 1 ,
\end{equation}
and the first correction reads as
\begin{eqnarray}
\widetilde{P}_c^{(1)} & = & 
\frac{\kappa}{(2\pi)^{3/2}\left( 1-{\rm e}^{-\kappa\widetilde{d}}\right)} \Bigg\{
F(\Delta) \left( \langle\widetilde{z}\rangle_0^2 
-\frac{\langle\widetilde{z}^2\rangle_0}{2} \right) \nonumber 
 + F(\Delta) {\rm e}^{-\kappa\widetilde{d}} \left[ 
\frac{\widetilde{d}^2-\langle(\widetilde{d}-\widetilde{z})^2\rangle_0}{2} 
- \langle(\widetilde{d}-\widetilde{z})\rangle_0 \left( \widetilde{d}
-\langle(\widetilde{d}-\widetilde{z})\rangle_0 \right) \right]
 \\ & &
+ 2\pi\frac{\partial\kappa(\eta,\Delta)}{\partial\eta}
\Bigg[ -\frac{\langle\widetilde{z}^2\rangle_0}{2}
+ {\rm e}^{-\kappa\widetilde{d}} \frac{\widetilde{d}^2-
\langle(\widetilde{d}-\widetilde{z})^2\rangle_0}{2} 
%\nonumber \\ & &
- \langle(\widetilde{d}-\widetilde{z})\rangle_0 \langle\widetilde{z}\rangle_0
+ {\rm e}^{-\kappa\widetilde{d}} \langle\widetilde{z}\rangle_0
\left( \widetilde{d} - \langle(\widetilde{d}-\widetilde{z})\rangle_0 \right) 
\Bigg] \Bigg\} .
\end{eqnarray}

%%%%%%%%%%%%%%%%%%%%%%%%%%%%%%%%%%%%%%%%%%%%%%%%%%%%%%%%%%%%%%%%%%%%%%%%%%%%%%

\section{Series representations of certain lattice 
functions} 
\label{app:C}
The function $F(\Delta)$ defined by Eq. (\ref{F}) corresponds to a special case of
$F(\Delta;{\bf q})$ introduced by expression \eqref{Fq}, since 
$F(\Delta)=F(\Delta,{\bf 0})$. This Fourier lattice sum can be written
as 
the series
\begin{eqnarray}
F(\Delta;{\bf q}) & = & - \frac{4}{3} \pi + \frac{4}{\sqrt{\pi}} 
\sum_{j=1}^{\infty} \left[ \cos(2\pi q_x j) z_{5/2}(0,j^2/\Delta) +
\cos(2\pi q_y j) z_{5/2}(0,j^2\Delta) \right] \nonumber \\ & &
+ \frac{8}{\sqrt{\pi}} \sum_{j,k=1}^{\infty} \cos(2\pi q_x j) \cos(2\pi q_y k) 
z_{5/2}(0,j^2/\Delta+k^2\Delta) \nonumber \\ & &
+ 2 \pi^{3/2} \sum_{j,k=-\infty}^{\infty}
z_{1/2}[0,(j-q_x)^2\Delta+(k-q_y)^2/\Delta] . \label{Fseries}
\end{eqnarray}
The function $G(\eta,\Delta;{\bf q})$ defined by Eq. (\ref{Gq}) is expressible 
as the series
\begin{eqnarray}
G(\eta,\Delta;{\bf q}) & = & \frac{8}{\sqrt{\pi}} \sum_{j,k=1}^{\infty}
\cos\left[ 2\pi q_x (j-1/2)\right] \cos\left[ 2\pi q_y (k-1/2)\right] 
z_{5/2}[0,(j-1/2)^2/\Delta+(k-1/2)^2\Delta+\eta^2] \nonumber \\ & &
+ 2 \pi^{3/2} \sum_{j,k=-\infty}^{\infty} (-1)^j (-1)^k
z_{1/2}[(\pi\eta)^2,(j-q_x)^2\Delta+(k-q_y)^2/\Delta] . \label{Gseries}
\end{eqnarray}


\begin{thebibliography}{10}

\bibitem{Palberg04} T. Palberg, M. Medebach, N. Garbow, M. Evers, 
A. Barreira Fontecha, H. Reiber, and E. Bartsch, 
J. Phys.: Condens. Matter \textbf{16}, S4039 (2004).

\bibitem{Attard96} Ph. Attard,
Adv. Chem. Phys. \textbf{92}, 1 (1996).

\bibitem{Hansen00} J.P. Hansen and H. L\"owen,
Annu. Rev. Phys. Chem. \textbf{51}, 209 (2000).

\bibitem{Levin02} Y. Levin,
Rep. Prog. Phys. \textbf{65}, 1577 (2002).

\bibitem{Messina09} R. Messina,
J. Phys.: Condens. Matter \textbf{21}, 113102 (2009).

\bibitem{Khan85} A. Khan, B. J\"onsson, and H. Wennerstr\"om,
J. Chem. Phys. \textbf{89}, 5180 (1985).

\bibitem{Kjellander88} R. Kjellander, S. Mar\v{c}elja, and J. P. Quirk,
J. Colloid Interface Sci. \textbf{126}, 194 (1988).

\bibitem{Bloomfield91} V. A. Bloomfield, 
Biopolymers \textbf{31}, 1471 (1991).

\bibitem{Rau92} D. C. Rau and A. Pargesian, 
Biophys. J. \textbf{61}, 246 (1992); ibid. \textbf{61}, 260 (1992).

\bibitem{Kekicheff93} P. K\'ekicheff, S. Mar\v{c}elja, T. J. Senden, and 
V. E. Shubin, J. Chem. Phys. \textbf{99}, 6098 (1993).

\bibitem{Dubois98} M. Dubois, T. Zemb, N. Fuller, R. P. Rand, 
and V. A. Pargesian, J. Chem. Phys. \textbf{108}, 7855 (1998).

\bibitem{Gulbrand84} L. Gulbrand, B. J\"onsson, H. Wennerstr\"om, and P. Linse,
J. Chem. Phys. \textbf{80}, 2221 (1984).

\bibitem{Kjellander84} R. Kjellander and S. Mar\v{c}elja,
Chem. Phys. Lett. \textbf{112}, 49 (1984).

\bibitem{Bratko86} D. Bratko, B. J\"onsson, and H. Wennerstr\"om,
Chem. Phys. Lett. \textbf{128}, 449 (1986).

\bibitem{Gronbech97} N. Gr{\o}nbech-Jensen, R. J. Mashl, R. F. Bruinsma,
and W. M. Gelbart, Phys. Rev. Lett. \textbf{78}, 2477 (1997). 

\bibitem{Linse99} P. Linse and V. Lobaskin,
Phys. Rev. Lett. \textbf{83}, 4208 (1999).

\bibitem{LinseLobaskin2000} P. Linse and V. Lobaskin,
J. Chem. Phys. \textbf{112}, 3917 (2000).

\bibitem{Bloomfield96} V. A. Bloomfield, 
Curr. Opin. Struct. Biol. \textbf{6}, 334 (1996).

\bibitem{May08}
S. May, A. Iglic, J. Rescic, S. Maset, and K. Bohinc,
J. Phys. Chem. B \textbf{112}, 1685 (2008).

\bibitem{Kim08}
Y. W. Kim, J. Yi, and P. A. Pincus, 
Phys. Rev. Lett. \textbf{101}, 208305 (2008).

\bibitem{LobaskinLinseJCP1999}
V. Lobaskin and P. Linse,
J. Chem. Phys. \textbf{111}, 4300 (1999).

\bibitem{Andelman06} D. Andelman, in
{\it Soft Condensed Matter Physics in Molecular and Cell Biology},
edited by W.C.K. Poon and D Andelman (Taylor \& Francis, New York, 2006). 

\bibitem{Attard88} Ph. Attard, D. J. Mitchell, and B. W. Ninham,
J. Chem Phys. \textbf{88}, 4987 (1988); \textbf{89}, 4358 (1988);
R. Podgornik, J. Phys. A \textbf{23}, 275 (1990);
R. R. Netz and H. Orland, Eur. Phys. J. E \textbf{1}, 203 (2000). 

\bibitem{Shklovskii99} B. I. Shklovskii,
Phys. Rev. E \textbf{60}, 5802 (1999);
Phys. Rev. Lett. \textbf{82}, 3268 (1999).

\bibitem{Chen06} Y. G. Chen and J. D. Weeks,
Proc. Natl. Acad. Sci. U. S. A. \textbf{103}, 7560 (2006);
J. M. Rodgers, C. Kaur, and Y. G. Chen, 
Phys. Rev. Lett. \textbf{97}, 097801 (2006).

\bibitem{Santos09} A. P. dos Santos, A. Diehl, and Y. Levin,
J. Chem. Phys. \textbf{130}, 124110 (2009).

\bibitem{BausHansen80}
M. Baus and J.-P. Hansen, Phys. Rep. \textbf{59}, 1 (1980).

\bibitem{Earnshaw1842} S. Earnshaw,
Trans. Cambridge Philos. Soc. \textbf{7}, 97 (1842).

\bibitem{Falko94} V.I. Falko,
Phys. Rev. B \textbf{49}, 7774 (1994).

\bibitem{Esfarjani95} K. Esfarjani and Y. Kawazoe,
J. Phys.: Condens. Matter \textbf{7} 7217 (1995).

\bibitem{Goldoni96} G. Goldoni and F.M. Peeters,
Phys. Rev. B \textbf{53}, 4591 (1996).

\bibitem{Schweigert99} I. V. Schweigert, V. A. Schweigert, and F. M. Peeters, 
Phys. Rev. Lett. \textbf{82}, 5293 (1999); Phys. Rev. B \textbf{60}, 14 665
(1999).

\bibitem{Weis01} J. J. Weis, D. Levesque, and S. Jorge,
Phys. Rev. B \textbf{63}, 045308 (2001).

\bibitem{Messina03} R. Messina and H. L\"owen, 
Phys. Rev. Lett. \textbf{91}, 146101 (2003); 
E. C. O\v{g}uz, R. Messina, and H. L\"owen, 
Europhys. Lett. \textbf{86}, 28002 (2009).

\bibitem{Lobaskin07} V. Lobaskin and R. R. Netz,
Europhys. Lett. \textbf{77}, 38003 (2007).

\bibitem{Samaj12a} L. \v{S}amaj and E. Trizac,
Europhys. Lett. \textbf{98}, 36004 (2012); 
Phys. Rev. B \textbf{85}, 205131 (2012).

\bibitem{Misra} R. D. Misra, Math. Proc. Cambridge Philos. Soc.
\textbf{36}, 173 (1940); 
M. Born and R. D. Misra, Math. Proc. Cambridge Philos. Soc.
\textbf{36}, 466 (1940). 

\bibitem{Moritz16}
M. Antlanger, G. Kahl, M. Mazars, L. \v{S}amaj, and  E. Trizac,  
Phys. Rev. Lett. \textbf{117}, 118002 (2016).

\bibitem{Samaj12b} L. \v{S}amaj and E. Trizac,
Contrib. Plasma Phys. \textbf{52}, 53 (2012);
Europhys. Lett. \textbf{100}, 56005 (2012). 

\bibitem{Rouzina96} I. Rouzina and V. A. Bloomfield,
J. Phys. Chem. \textbf{100}, 9977 (1996).

\bibitem{Perel99} V. I. Perel and B. I. Shklovskii, 
Physica A \textbf{274}, 446 (1999).

\bibitem{Moreira00} A.G. Moreira and R.R. Netz:
Europhys. Lett. \textbf{52}, 705 (2000); 
Phys. Rev. Lett. \textbf{87}, 078301 (2001).

\bibitem{Netz01} R.R. Netz:
Eur. Phys. J. E \textbf{5}, 557 (2001).

\bibitem{Moreira02} A.G. Moreira and R.R. Netz:
Eur. Phys. J. E \textbf{8}, 33 (2002).

\bibitem{Santangelo06} C. D. Santangelo,
Phys. Rev. E \textbf{73}, 041512 (2006).

\bibitem{Samaj11} L. \v{S}amaj and E. Trizac,
Phys. Rev. Lett. \textbf{106}, 078301 (2011);
Phys. Rev. E \textbf{84}, 041401 (2011).

\bibitem{Kanduc07} M. Kandu\v{c} and R. Podgornik,
Eur. Phys. J. E \textbf{23}, 265 (2007);
Y. S. Jho, M. Kandu\v{c}, A. Naji, R. Podgornik, M. W. Kim, and P. A. Pincus,
Phys. Rev. Lett. \textbf{101}, 188101 (2008).

\bibitem{Kanduc08} M. Kandu\v{c}, M. Trulsson, A. Naji, Y. Burak, J. Forsman, 
and R. Podgornik, Phys. Rev. E \textbf{78}, 061105 (2008). 

\bibitem{Paillusson11} F. Paillusson and E. Trizac,
Phys. Rev. E \textbf{84}, 011407 (2011). 

\bibitem{Kanduc10} M. Kandu\v{c}, A. Naji, J. Forsman, and R. Podgornik,
J. Chem. Phys. \textbf{132}, 124701 (2010); 
Phys. Rev. E \textbf{84}, 011502 (2011). 

\bibitem{Burak04} Y. Burak, D. Andelman, and H. Orland,
Phys. Rev. E \textbf{70}, 016102 (2004).

\bibitem{Nordholm84} S. Nordholm,
Chem. Phys. Lett. \textbf{105}, 302 (1984).

\bibitem{Hatlo10} M. M. Hatlo and L. Lue,
EPL \textbf{89}, 25002 (2010).

\bibitem{Samaj16} L. \v{S}amaj, A. P. dos Santos, Y. Levin, and E. Trizac,
Soft Matter \textbf{12}, 8768 (2016).

\bibitem{Ivan} I. Palia, M. Trulsson, L. \v{S}amaj, and E. Trizac,
arXiv:1803.00359, submitted (2018).  

\bibitem{Berkowitz} I.-C. Yeh and M. L. Berkowitz, 
J. Chem. Phys. \textbf{111}, 3155 (1999).

\bibitem{Mazars} M. Mazars, J.-M. Caillol, J.-J. Weis, and D. Levesque, 
Condens. Matter Phys. \textbf{4}, 697 (2001).

\bibitem{Serr} H. Boroudjerdi, Y.-W. Kim, A. Naji, R. R. Netz, 
X. Schlagberger, and A. Serr, 
Phys. Rep. \textbf{416}, 129 (2005).

\bibitem{Grimes79} C. C. Grimes and G. Adams,
Phys. Rev. Lett. \textbf{42}, 795 (1979). 

\bibitem{Morf79} R. H. Morf,
Phys. Rev. Lett. \textbf{43}, 931 (1979).

\bibitem{contact} D. Henderson and L. Blum, J. Chem. Phys. \textbf{69}, 
5441 (1978); D. Henderson, L. Blum, and J.L. Lebowitz, 
J. Electroanal. Chem. \textbf{102}, 315 (1979);
S.L. Carnie, D.Y.C. Chan, J. Chem. Phys. \textbf{74}, 1293 (1981);
H. Wennerstr\"om, B. J\"onsson, and P. Linse,
J. Chem. Phys. \textbf{76}, 4665 (1982).

\bibitem{MaTT15}
J.-P. Mallarino,   G. T\'ellez,   E. Trizac,
Mol. Phys. \textbf{113}, 2409 (2015).

\bibitem{HansenMcDonald}
J.-P. Hansen and I. R. McDonald, {\it Theory of Simple Liquids}, Academic Press,
Amsterdam (2007).

\bibitem{Gradshteyn} I. S. Gradshteyn and I. M. Ryzhik,
Table of Integrals, Series, and Products, 6th ed.
(Academic, London, 2000).

\bibitem{Travenec15} I. Trav\v{e}nec and L. \v{S}amaj,
Phys. Rev. E \textbf{92}, 022306 (2015). 


\end{thebibliography}
\end{document}